\documentclass[prb,twocolumn,showpacs]{revtex4}

\usepackage{graphicx}
\usepackage{bm}
\usepackage{times}
\RequirePackage{amsmath2000}

\newcommand{\smeq}{\! = \!}

\newcommand{\smpl}{\! + \!}
\newcommand{\smmi}{\! - \!}

\newcommand{\Vtf}{V_{\text{TF}}}

\newcommand{\br}{{\bm{r}}}
\newcommand{\Vsc}{V_{\text{sc}}}
\newcommand{\e}{\epsilon}
\newcommand{\ve}{\varepsilon}
\newcommand{\kf}{k_{\text{F}}}
\newcommand{\Ef}{E_{\text{F}}}
\newcommand{\Eth}{E_{\text{th}}}

\newcommand{\ag}{{\mathcal{N}}}
\newcommand{\kt}{k_{\text{B}}T}
\newcommand{\Egs}{E_{\text GS}}
\newcommand{\be}{\begin{equation}}
\newcommand{\ee}{\end{equation}}
\newcommand{\bea}{\begin{eqnarray}}
\newcommand{\eea}{\end{eqnarray}}
\newcommand{\Ha}{{\hat H}}
\newcommand{\Hint}{{\hat H}_{\text{int}}}
\newcommand{\up}{\uparrow}
\newcommand{\dn}{\downarrow}

\begin{document}
\title{Spin and \textit{e-e} interactions in quantum dots: Leading order 
corrections to universality and temperature effects}
\author{Gonzalo Usaj}
\author{Harold U. Baranger}
\affiliation{Department of Physics, Duke University, P. O. Box 90305, Durham NC 
27708-0305}
\date{\today}

   \begin{abstract}
We study the statistics of the spacing between Coulomb blockade conductance
peaks in quantum dots with large dimensionless conductance $g$.  Our starting
point is the ``universal Hamiltonian''---valid in the $g\!\rightarrow\!\infty$
limit---which includes the charging energy, the single-electron energies
(described by random matrix theory), and the average exchange interaction. We
then calculate the magnitude of the most relevant finite $g$ corrections,
namely, the effect of surface charge, the ``gate'' effect, and the fluctuation
of the residual \textit{e-e} interaction.  The resulting zero-temperature peak
spacing distribution has corrections of order $\Delta/\sqrt{g}$.  For typical
values of the \textit{e-e} interaction ($r_{s}\!\sim\!1$) and simple geometries,
theory does indeed predict an \textit{asymmetric} distribution with a significant
\textit{even/odd} effect. The width of the distribution is of order $0.3\Delta$,
and its dominant feature is a large peak for the odd case, reminiscent of the
$\delta$-function in the $g\!\rightarrow\!\infty$ limit. We consider finite
temperature effects next. Only after their inclusion is good agreement with
the experimental results obtained. Even relatively low temperature causes
\textit{large} modifications in the peak spacing distribution: (a) its peak is
dominated by the \textit{even} distribution at $\kt\!\sim\!0.3\Delta$ (at lower $T$
a double peak appears); (b) it becomes more symmetric; (c) the even/odd effect is
considerably weaker; (d) the $\delta$-function is completely washed-out; and (e)
fluctuation of the coupling to the leads becomes relevant.  Experiments aimed at
observing the $T \!=\!0$ peak spacing distribution should therefore be done at
$T\!<\!0.1\Delta/k_{\text{B}}$ for typical values of the \textit{e-e} interaction.
   \end{abstract}

\maketitle 
%%%%%%%%%%%%%%%%%%%%%%%%%%%%%%%%%%%%%%%%%%%%%%%%%%%%%%%%%%%%%%%%%%%%%%%%%
\section{Introduction}
%%%%%%%%%%%%%%%%%%%%%%%%%%%%%%%%%%%%%%%%%%%%%%%%%%%%%%%%%%%%%%%%%%%%%%%%%

The Coulomb Blockade (CB) of electron tunneling is
one of the most studied effects in quantum dots (QDs).
\cite{GrabertD92,Kastner92,MesoTran97,KouwenetalRev97,Alhassid00,AleinerBG02}
It allows one to probe quantum interference effects in both the wavefunction
and the energy of interacting electrons.  The main way in which the latter
has been probed is through the spacing between adjacent CB conductance peaks.
A satisfactory explanation for the observed CB peak spacing distribution
(PSD) has, however, remained elusive. In this paper we first focus on the $T
\!=\! 0$ PSD and find its shape for quantum dots containing a few hundred
electrons.  We then turn to the effect of temperature, showing that it is
surprisingly large.  In the end reasonable quantitative agreement between
theory and experiments is obtained.

The CB effect occurs when the thermal energy $\kt$ is smaller than the charging
energy $E_{C}\smeq e^2/2C$ required to add an electron to the QD---$C$ is the total
capacitance of the QD. In that case, electron transport through the QD is blocked by
energetics, fixing the number of electrons $N$ in the QD. By sweeping the voltage
$V_g^{}$ of a capacitively coupled gate, this CB can be overcome at a particular
value $V_g^{N}$ where the transition $N\rightarrow\!N+1$ occurs. The conductance
$G(V_{g}^{})$ shows then a series of sharp peaks as a function of $V_g^{}$
as additional electrons are added to the QD.  At sufficiently low temperature,
only the ground state (GS) contributes significantly to the conductance peak. The
position of the CB peak is then proportional to the change in the GS energy of
the QD upon adding one electron.\cite{KouwenetalRev97}

The simplest model used for the description of this phenomenon assumes a constant
\textit{e-e} interaction---hence the name constant interaction (CI) model.
The single-particle part of the Hamiltonian is described by random matrix theory
(RMT) under the assumption that the single-particle classical dynamics is
chaotic (or diffusive). As a result, the fluctuation of both the conductance
peak height and the CB peak spacing are given by single-particle RMT statistics.
Despite the success of this model in explaining
the former\cite{JalabertSA92}---once thermal\cite{AlhassidGS98,VallejosLM99}
and periodic orbit effects\cite{NarimanovCBT99,Kaplan00,NarimanovBCT01}
are included---it fails drastically in describing the observed PSD.
\cite{SivanBAPAB96,SimmelHW97,PatelCSHMDHCG98,PatelSMGASDH98,SimmelAWKK99,LuscherHEWB01}

Within the CI model, electrons fill the states of the QD in an ``up-down" scheme due
to the spin degeneracy. This implies a strong  even/odd effect on the PSD. However,
none of the experiments to date have shown such an effect---though a weak even/odd
effect was observed\cite{LuscherHEWB01,OngBHPM01}---suggesting that spin plays
a more active role. Furthermore, the observed PSD presents a Gaussian-like shape
(with broader non-Gaussian tails), which contradicts the expected Wigner-Dyson
distribution from random matrix theory.  Finally, the magnitude of the width
of the PSD was questioned. Early experiments\cite{SivanBAPAB96,SimmelHW97}
found it scaled with $E_C$, which is much bigger than the predicted value,
$\sim\!\Delta$, the single-particle mean level spacing.  More recent
experiments,\cite{PatelCSHMDHCG98,SimmelAWKK99,LuscherHEWB01} however, showed
that it is indeed of order $\Delta$.

The search for an explanation to these
discrepancies triggered several theoretical works over the last years.
\cite{SivanBAPAB96,PrusAASB96,BlanterMM97,BerkovitsA97,Berkovits98,VallejosLM98,BrouwerOH99,BarangerUG00,KurlandAA00,JacquodS00,UllmoB01,UsajB01_RC,OregBWH01}
Fueled by the earlier experiments,\cite{SivanBAPAB96,SimmelHW97} it was
suggested\cite{SivanBAPAB96,PrusAASB96,BerkovitsA97,Berkovits98} that GS  fluctuations were
dominated by the \textit{e-e} interaction itself. Therefore, a completely different
approach---involving non-perturbative methods such as self-consistent Hartree-Fock
or exact diagonalization---was required. On the other hand, based on the
fact that a typical QD contains a large number of electrons, $N\gg1$, it was argued
\cite{BlanterMM97,AleinerG98,VallejosLM98,BrouwerOH99,BarangerUG00,KurlandAA00,JacquodS00,UllmoB01,UsajB01_RC,OregBWH01}
that they should be described as ``good'' metals. This implies that the
residual \textit{e-e} interactions (i.e., those beyond $E_C$) are weak and can
be added to the CI model perturbatively. We shall take the latter approach
and show that it provides a good description of the experimental data.
\cite{PatelCSHMDHCG98,OngBHPM01}

The small parameter in this perturbative approach is $1/g$ with 
$g\smeq\Eth/\Delta$ the dimensionless conductance and $\Eth$ the Thouless energy 
(approximately $\hbar$
 times the inverse time of flight). The condition for the QD to be a good 
conductor is $g\!\propto\!\sqrt{N}\gg1$.\cite{AleinerG98,AleinerBG02} 
Interaction corrections are classified by their order in $\Delta/g$ and 
successively added to the CI Hamiltonian. It then becomes clear why the CI model 
is wrong: there is a 
zero-order correction---i.e. a correction of order $\Delta$---namely, the 
\textit{average} exchange 
interaction.\cite{BrouwerOH99,BarangerUG00,KurlandAA00,UllmoB01} 
Although this is a small correction to the total energy of the QD---and so the 
perturbative approach is justified---it is crucial for properties, like the CB 
peak spacing, that 
are sensitive to single levels in the QD.
The zero-order Hamiltonian---hereafter called the constant exchange and 
interaction (CEI) model---is given by\cite{KurlandAA00,AleinerG98,AleinerBG02} 
\be
\Ha_{\text{CEI}}^{}\smeq\sum_{\alpha,\sigma}{\ve_{\alpha}}\,\hat{n}_{\alpha,\sigma}
                              \smpl E_{C}\,(\hat{n}-\ag)^2\smmi J_{S}\;{\vec S}^2
\label{CEI}
\ee
where $\{\ve_{\alpha}\}$ are the single-electron energies, $\ag\smeq
C_g V_g/e$ describes the capacitive coupling to the control gate, $C_g$
is the dot-gate capacitance, ${\vec S}$ is the total spin operator, and
$J_{S}$ is the exchange constant.  The difference between the CEI and CI
models is the additional term proportional to ${\vec S}^2$.  Because it
sometimes leads to a GS with $S\!\ge\!1$, the simple ``up-down" filling
scheme breaks down.\cite{BrouwerOH99,BarangerUG00,KurlandAA00,KurlandBA01} The
corresponding PSD is \textit{completely} different from the CI model result (see
Ref. \onlinecite{UllmoB01} for a plot).  In fact, a GS with $S\smeq1$  has been
experimentally observed very recently.\cite{FolkMBKAA01} The PSD resulting from
the CEI model is still, however, in poor agreement with the data.

Most of the work so far has concentrated on the calculation of higher
order corrections to the Hamiltonian. The most important ones are:
(1) the ``scrambling" of the spectrum when adding an electron to the
QD;\cite{BlanterMM97} (2) the fluctuation of the diagonal matrix element of
the \textit{e-e} interaction;\cite{UllmoB01} and finally, not related to the
\textit{e-e} interaction, (3) the change in the single-electron energies when
the gate voltage is swept.\cite{VallejosLM98} Surprisingly, although these
corrections have been discussed previously in the literature,  an explicit
calculation of the PSD including all of them has not been done---note however
that Ref. \onlinecite{UllmoB01} included the first two.  Here we present
results that include all three effects and show that the scrambling and
gate effects, though dominant, are much smaller than usually assumed in the
literature.\cite{VallejosLM98,AlhassidM99}  We also show that the fluctuation of
the off-diagonal matrix elements\cite{JacquodS00,JacquodS01} introduces a small
correction---in the regime relevant for the experiments---and can be disregarded in
the calculation of the PSD.  Despite the substantial improvement these corrections
introduce, the disagreement with the experimental results persists.

Very recently,\cite{UsajB01_RC} we pointed out that there is a simple effect
that has not been taken into account: finite temperature---we should mention
though that temperature effects have been discussed in terms of spinless
particles.\cite{PatelSMGASDH98,AlhassidM99} In our previous paper,\cite{UsajB01_RC}
we show that in the CEI model the temperature effects are more important than in the
CI model and that they become significant even at $\kt\!\sim\!0.1\Delta$. Since most
experiments were done in the regime $\kt\!\sim\!0.3$-$0.5\Delta$---an exception
is Ref.\onlinecite{LuscherHEWB01}---our results are crucial for interpreting
the experimental data.  Here, we present those results in more detail and extend
them. We show that, because of temperature, the fluctuation of the coupling to the
leads strongly affects the PSD.  It is not until temperature is introduced that
good agreement with the experimental results is obtained. Furthermore, temperature
introduces the biggest correction in the regime where most experiments were done
so far and constitutes the main cause of smoothing of the PSD.  Consequently,
lower temperature experiments are required in order to observe the actual ground
state PSD.

The rest of the paper is organized as follows: in Section \ref{CEImodel} we
review the arguments that lead to the CEI model.  The leading order corrections
to this model are introduced in Section \ref{sectionleading}. We calculate
the contribution of the off-diagonal terms of the \textit{e-e} interaction
and the magnitude of the scrambling effect in Section \ref{sectionoff} and
\ref{evalvarchi} respectively. Numerical results for the $T\smeq0$ PSD, including
all the corrections, are presented in Section \ref{results}. We introduce the
effect of finite temperature in Section \ref{temp}.  Finally, we conclude in
Section \ref{conclu}.

%
%%%%%%%%%%%%%%%%%%%%%%%%%%%%%%%%%%%%%%%%%%%%%%%%%%%%%%%%%%%%%%%%%%%%%%%%%
\section{The CEI model\label{CEImodel}}
%%%%%%%%%%%%%%%%%%%%%%%%%%%%%%%%%%%%%%%%%%%%%%%%%%%%%%%%%%%%%%%%%%%%%%%%%
%
At low temperature and low bias, only a few energy levels around the Fermi
energy ($\Ef$) are involved in the transport process. Consequently, an effective
Hamiltonian $\Ha_{\text{QD}}$ capable of describing the QD in that energy window
is all that is needed. When considering single-particle properties of chaotic
(or diffusive) QDs, it is well-known that $\Ha_{\text{QD}}$  can be described
by random matrix theory (RMT) provided that $g\!\gg\!1$. This approach is valid
within an energy window up to the Thouless energy $\Eth$.  The single-particle
Hamiltonian is then ``universal'', i.e. it only depends on the symmetry of the
problem---broken time-reversal symmetry is assumed throughout the paper---and, of
course, the energy scale $\Delta$. This approach has proved to be quite successful
for understanding the role of mesoscopic fluctuations in transport properties of QD
(see Refs. \onlinecite{Alhassid00} and  \onlinecite{Beenakker97} for reviews).

On the other hand, the treatment of the $e$-$e$ interaction is more subtle.
A proper description requires taking into account the screening of the bare
Coulomb interaction provided by the electrons beyond $\Eth$.  If the interactions
are not so strong, i.e. if the gas parameter $r_s$ is small, the screening can
be calculated using the random phase approximation. In that case, the screened
potential reads\cite{BlanterMM97,AleinerBG02}
\be
\Vsc(\br_1,\br_2)\smeq\frac{e^2}{C}\smpl\Vtf(\br_1,\br_2)\smpl 
V(\br_1)\Delta\smpl V(\br_2)\Delta
\label{scpotential}
\ee
where $C$ is the capacitance of the QD, $\Vtf(\br)$ is the
Thomas-Fermi screened potential and $V(\br)\Delta$ is a finite-size screened
potential\cite{BlanterMM97}---its specific form and origin will be discussed in
Section \ref{evalvarchi}. Here and throughout this paper we consider two-dimensional
(2D) quantum dots. The last three terms in Eq. (\ref{scpotential}) are of order
$\Delta$ and so much smaller than the first term.  This leads to the following
Coulomb interaction Hamiltonian
\begin{eqnarray}
\nonumber
\Ha_{\text{int}}&\smeq& 
E_{C}\,(\hat{n}^2-\hat{n})\smpl(\hat{n}\smmi1)\sum_{\alpha,\beta,\sigma}\,{c^{\dagger}_{\alpha,\sigma}c^{}_{\beta,\sigma}
                                      \,{\cal X}^{}_{\alpha,\beta}}\\
                        &         & +\frac{1}{2}
               \sum_{\alpha,\beta,\gamma,\delta}{H_{\alpha,\beta,\gamma,\delta}^{}\;
               \; c^{\dagger}_{\delta,\sigma}c^{\dagger}_{\gamma,\sigma'}
               c^{}_{\beta,\sigma'}c^{}_{\alpha,\sigma}}
\label{intH}
\end{eqnarray}
where 
\bea
\nonumber
H_{\alpha,\beta,\gamma,\delta}^{}&\smeq&
   \int\text{d}\br_1\text{d}\br_2\Psi^{*}_{\delta}(\br_1)\Psi^{*}_{\gamma}(\br_2)\\
   &&\qquad\times  
\Vtf(\br_1\smmi\br_2)\Psi^{}_{\beta}(\br_2)\Psi^{}_{\alpha}(\br_1)\, ,
\label{matrixelement} 
\eea
$\mathcal{X}_{\alpha,\beta}\smeq\Delta\int\text{d}\br V(\br) 
\Psi^{*}_{\alpha}(\br)\Psi^{}_{\beta}(\br)$
%\label{defchi}
% 
and $\Psi^{}_{\alpha}(\br)$ is the eigenfunction of the single-electron Hamiltonian
with eigenvalue $\ve_{\alpha}$.  Fluctuations of the wavefunctions cause both
$H_{\alpha,\beta,\gamma,\delta}^{}$ and $\mathcal{X}_{\alpha,\beta}$ to fluctuate.
However, since both matrix elements are defined as integrals over the QD's area,
$A$, one could expect the main contribution to come from their mean value due to
self-averaging. It turns out that this is indeed the case and that the parameter
that controls the relative importance of the fluctuations is $1/g$ (or equivalently
$1/\kf \sqrt{A}$).  This suggests we can expand $\Ha_{\text{int}}$ in powers of
$1/g$ and keep only the dominant terms.\cite{AleinerG98,KurlandAA00,AleinerBG02}

The zeroth-order term ($g\!\rightarrow\!\infty$) in this expansion corresponds to 
taking the mean value of the matrix elements,
\be
\langle H_{\alpha,\beta,\gamma,\delta}^{}\rangle^{(0)}\smeq J'\, 
\delta_{\alpha,\delta}\delta_{\beta,\gamma}\smpl J_{S}\, 
\delta_{\alpha,\gamma}\delta_{\beta,\delta}
\label{mean}
\ee
where $J'\smeq A^{\smmi2}\int\text{ d}\br_1\text{ d}\br_2\, 
\Vtf(\br_1-\br_2)\!\simeq\!\Delta/2$ and
\be
J_{S}\smeq\frac{1}{A}\int\text{ d}\br\, \Vtf(\br)\, J_0^2(\kf r)
\label{JS}
\ee
is the exchange constant, $J_0(x)$ is a Bessel function (2D case) and 
${\bf k}_{\text {F}}$ is the Fermi wavevector.  We shall see in Section
\ref{evalvarchi} that 
$\langle \mathcal{X}_{\alpha,\beta}\rangle\smeq\Delta/2\, \delta_{\alpha,\beta}$.
Introducing these mean values in Eq. (\ref{intH}) we obtain the ``universal" part 
of the interaction Hamiltonian\cite{KurlandAA00,AleinerBG02} 
\be
\Ha_{\text{int}}^{(0)}\smeq E_{C}\,\hat{n}^2\smmi J_{S}\;{\vec S}^2
\label{uH}
\ee
where we have dropped terms linear in $\hat{n}$ and redefine
$E_C$ to include all the terms proportional to $\hat{n}^2$.  Here, ${\vec
S}\smeq\sum_{\alpha,\sigma,\sigma'}{c^{\dagger}_{\alpha,\sigma}\vec{\sigma}_{\sigma,\sigma'}c^{}_{\alpha,\sigma'}}$
is the total spin operator of the QD. Eq. (\ref{uH}) is the most general form
of the interaction Hamiltonian compatible with RMT.\cite{KurlandAA00} Adding the
coupling to the control gate gives the CEI model Hamiltonian [Eq. (\ref{CEI})].

Using the explicit expression for $\Vtf(\br)$,\cite{UllmoB01}
%
%\be
%\Vtf(\br)\smeq\frac{1}{2\pi}\int\text{d}\bm{q} \frac{\Delta A/2}{q(\kf r_{s}\sqrt{2})^{\smmi1}\smpl1}e^{\smmi i \bm{q}.\br}
%\ee
%
 we can write $J_S$ in terms of the gas parameter $r_s$, 
\be
J_S\smeq\frac{r_{s}\Delta}{\pi}\frac{\text{Arc}\!\sec{(r_{s}/\sqrt{2})}}{\sqrt{r_{s}^{2}\smmi2}}\qquad .
\ee
Note that $J_S\!\leq\!\Delta/2$ is a fixed quantity. Fluctuations in the spectrum 
of $\Ha_\text{CEI}$ arise only from 
$\{\ve_{\alpha}\}$. This is a key point for understanding its GS: while in the CI 
model the levels are filled in an ``up-down" scheme---which leads to a bimodal 
PSD---in the CEI model it is energetically favorable to promote an electron to a 
higher level and gain exchange energy whenever the spacing between the top two 
single-particle levels is smaller than $2 J_S$ ($N$ even). This is why the PSD is 
very different from the CI model result.

%
%%%%%%%%%%%%%%%%%%%%%%%%%%%%%%%%%%%%%%%%%%%%%%%%%%%%%%%%%%%%%%%%%%%%%%%%%
\section{Leading order corrections to universality \label{sectionleading}}
%%%%%%%%%%%%%%%%%%%%%%%%%%%%%%%%%%%%%%%%%%%%%%%%%%%%%%%%%%%%%%%%%%%%%%%%%
%
In the previous section we considered the simplest model for $\Ha_{\text{QD}}$,
which only takes the universal part of the residual interactions into account.
Here, we include the next order corrections: (1) the ``scrambling" of the spectrum
when adding an electron to the QD;\cite{BlanterMM97} (2) the fluctuation of the
diagonal part of the $e$-$e$ interaction;\cite{UllmoB01} and (3) the change in
the single-electron energies when the gate voltage is swept.\cite{VallejosLM98}
Although (2) is a correction of order $\Delta/g$ to the Hamiltonian, all three
effects lead to corrections of order $\Delta/\sqrt{ g}$ to the spacing.  We discuss
now each of them in detail.

%%%%%%%%%%%%%%%%%%%%%%%%%%%%%%%%%%%%%%%%%%%%%%%%%%%%%%%%%%%%%%%%%%%%%%%%%
\subsection{Scrambling}
%%%%%%%%%%%%%%%%%%%%%%%%%%%%%%%%%%%%%%%%%%%%%%%%%%%%%%%%%%%%%%%%%%%%%%%%%
The scrambling effect\cite{BlanterMM97} is caused by the presence of the 
potential $V(\br)\Delta$ in Eq. (\ref{scpotential}), which 
leads to the second term in Eq. (\ref{intH}). Its physical origin is quite 
simple. When an electron is added to the QD, the other 
electrons arrange themselves to screen the extra charge. 
That means that a charge $-e/\kappa$ is pull out to the boundaries of the QD. 
This extra charge creates an additional potential, 
$V(\br)\Delta$, 
for the electrons inside the QD. While in $3$-D materials the charge is confined 
to a small region near the surface, in $2$-D it is inhomogeneously 
distributed over the whole area of the QD. 

It is straightforward to show that 
$\langle\mathcal{X}_{\alpha,\beta}\rangle\smeq\bar{V}\Delta\,\delta_{\alpha,\beta}$ 
where 
$\bar{V}\smeq A^{\smmi1}\int \text{d}\br V(\br)$. The mean value thus does not 
introduce any correction to the CEI model---it should be added to $E_C$.
The correction comes only from the fluctuations of $\mathcal{X}_{\alpha,\beta}$. 
In particular, the main contribution (to the spacing) arises from 
$\mathcal{X}_{\alpha,\alpha}$, with $\alpha$ the top level.\cite{BlanterMM97,UllmoB01} Its 
variance is given by,  
\be
\text{var}(\mathcal{X}_{\alpha,\alpha})\smeq
    \Delta^2\int\text{d}\br_1\text{d}\br_2 
\tilde{V}(\br_1)\tilde{V}(\br_2)\langle\left|\Psi^{*}_{\alpha}(\br_1)\Psi^{}_{\alpha}(\br_2)\right|^2\rangle. 
\label{varchi0}
\ee
with $\tilde{V}(\br)\smeq V(\br)\smmi\bar{V}$. 

With corrections of order $1/g$ included, the wavefunction correlation appearing 
in Eq. (\ref{varchi0}) is given 
by\cite{BlanterMM98,Mirlin00,BlanterMM01,GornyiM01}
\bea
\nonumber
A^2\langle\left|\Psi^{*}_{\alpha}(\br_1)\Psi^{}_{\alpha}(\br_2)\right|^2\rangle&\smeq&
                1+k(\br_1,\br_2)-k(\br_1)-k(\br_2)\\
\nonumber
& & +k+\Pi_{\text{B}}(\br_1,\br_2)\\
\label{correlation}
\eea
with $k(\br_1,\br_2)\smeq J_0^2(\kf\left|\br_1-\br_2\right|)$, $k(\br)\smeq
A^{\smmi1}\int \text{d}\br_1 k(\br,\br_1)$, and $k\smeq A^{\smmi2}\int \text{d}\br_1
\text{d}\br_2 k(\br_1,\br_2)$.  $\Pi_{\text{B}}(\br_1,\br_2)$ is a classical
propagator that contains the contributions of the trajectories that reach the
boundary of the QD---it is therefore geometry dependent. The latter satisfies
$\int \text{d}\br_i \Pi_{\text{B}}(\br_1,\br_2)\smeq0$ with $i\smeq1$ or $2$.
The second term in Eq. (\ref{correlation}) corresponds to Berry's result:
on scales smaller than the system size $\sqrt{A}$, the correlation of chaotic
wavefunctions is giving by a random superposition of plane waves.\cite{Berry77}
The other terms involving $k(\br_1,\br_2)$ properly account for the normalization
of the wavefunctions.  \cite{GornyiM01}

After substituting (\ref{correlation}) in (\ref{varchi0}) we get 
\bea
\nonumber
\text{var}(\mathcal{X}_{\alpha,\alpha})&\smeq&\frac{\Delta^2}{A^2}\int\text{d}\br_1\text{d}\br_2 
\tilde{V}(\br_1)\tilde{V}(\br_2) k(\br_1,\br_2) \\
\nonumber
 &&\smpl\frac{\Delta^2}{A^2}\int\text{d}\br_1\text{d}\br_2 \tilde{V}(\br_1) 
\tilde{V}(\br_2)\Pi_{\text{B}}(\br_1,\br_2).\\
\label{varchi}
\eea
For quantitative evaluation we consider the case of a ballistic circular disc 
with diffusive boundary conditions. For this system\cite{BlanterMM01} 
\be
\Pi_{\text{B}}(\br_1,\br_2)\smeq\frac{1}{4\pi\kf 
R}\sum_{q=1}^\infty{\frac{4q^2\smmi1}{4q^2}\left(\frac{r_1 
r_2}{R^2}\right)^q\cos{q(\theta_1\smmi\theta_2)}}
\label{PI}
\ee
where $\br_i\smeq r_i\,(\cos\theta_i,\sin\theta_i)$. Thus the last term in
(\ref{varchi}) yields $0$ exactly since the potential $V(\br)$ is isotropic 
(in the isolated dot case, see Section \ref{evalvarchi}).
For a general geometry, the form of $\Pi_{\text{B}}$ is not known, and there
is no reason \textit{a priori} to expect such a cancellation. We assume that the
contribution of this term is of the same order as the first and that our final
result for the $\text{var}(\mathcal{X}_{\alpha,\alpha})$ is correct up to a
factor $2$.  \footnote{Note that in the case of a diffusive QD, this term gives
the main contribution to the variance (see Appendix A), which turns out to be of
the same order as the first term in the ballistic case. It might also be important in
systems with strong semiclassical effects.}

Since $k(\br_1,\br_2)\!\approx\!1/\pi\kf|\br_1\smmi\br_2|$, a simple dimensional 
analysis of the first term in 
Eq. (\ref{varchi}) shows that it is proportional to $\Delta^2/\kf \sqrt{A}$ so 
that\cite{BlanterMM97,AleinerBG02}  
\be
\text{var}(\mathcal{X}_{\alpha,\alpha})\smeq b_{00}\frac{\Delta^2}{g}.
\label{varscfinal}
\ee
The same result is valid for $\text{var}(\mathcal{X}_{\alpha,\beta})$. Again
using the circular disc with diffusive boundaries for
quantitative estimation we have $\Eth\smeq\hbar\gamma_1
v_{\text{F}}/R$ so that $g\smeq\gamma_1\kf\sqrt{A}/2\sqrt{\pi}$ with
$\gamma_1\smeq0.38$.\cite{BlanterMM97,BlanterMM01} The approximate value of the
geometry-dependent coefficient $b_{00}$ is calculated in Section \ref{evalvarchi}.

%%%%%%%%%%%%%%%%%%%%%%%%%%%%%%%%%%%%%%%%%%%%%%%%%%%%%%%%%%%%%%%%%%%%%%%%%
\subsection{Diagonal matrix elements\label{diag}}
%%%%%%%%%%%%%%%%%%%%%%%%%%%%%%%%%%%%%%%%%%%%%%%%%%%%%%%%%%%%%%%%%%%%%%%%%
Recently, Ref. \onlinecite{UllmoB01} has shown that the fluctuation of the 
diagonal terms of $\Ha_{\text{int}}$ leads to a correction to the spacing of the 
same order as the scrambling, despite the fact that the variance of the diagonal 
matrix elements is of order $\Delta^{2}/g^{2}$. The reason is that the 
correction due to these terms involves a sum over $\approx g$ levels.
To see this, let us first calculate the variance of the diagonal matrix elements. 
In the  zero-range approximation, where the short-range screened potential is 
approximated by a $\delta$-function, $\Vtf(\br) \!\approx\! \delta(\br)\Delta A/2$, 
it is given by
\be
\text{var}(M_{\alpha,\beta})\smeq\frac{\Delta^2}{4A^2}\int\text{d}\br_1\text{d}\br_2\, 
\left[\tilde{k}(\br_1\smmi\br_2)\smpl\Pi_{\text{B}}(\br_1,\br_2)\right]^{2}\\
\label{varM}
\ee
where $M_{\alpha,\beta}\smeq H_{\alpha,\beta,\beta,\alpha}^{}\smmi\langle 
H_{\alpha,\beta,\beta,\alpha}\rangle^{(0)}$ and 
$\tilde{k}(\br_1\smmi\br_2)\smeq k(\br_1\smmi\br_2)\smmi k(\br_1)\smmi 
k(\br_2)\smpl k$. Using the full expression for $\Vtf$ leads to similar 
numerical results\cite{UllmoB01}---in that case it is important to keep the 
correlation between the direct and the exchange terms. 
The dominant contribution in Eq (\ref{varM}) comes from 
$[k(\br_1\smmi\br_2)]^{2}$---the other terms are of the same order in $1/g$ but 
numerically much smaller---so
\be
\text{var}(M_{\alpha,\beta})\smeq\frac{3\Delta^2}{4\pi} \frac{\ln(4\kf 
\sqrt{A})}{(\kf \sqrt{A})^{2}}
                                           \!\sim\! \frac{\ln{g}}{g^{2}}\Delta^2.
\label{varM1}
\ee
For the double-diagonal matrix element, 
$\text{var}(M_{\alpha,\alpha})\smeq4\text{var}(M_{\alpha,\beta})$.\cite{UllmoB01} 
Let us now consider the contribution of the fluctuation of the diagonal terms to 
the spacing corresponding to the transition 
$\frac{1}{2}\!\rightarrow\!0\!\rightarrow\!\frac{1}{2}$,  
\be
s_{\text{diag}}\smeq \sum_{\beta=\frac{N}{2}\smmi g}^{\frac{N}{2}}
{(M_{\frac{N}{2}\smpl1,\beta}\smmi M_{\frac{N}{2},\beta})}.
\ee
Neglecting the correlation between the different matrix
elements, \footnote{One should note, however, that the covariance between the matrix
elements $M_{\alpha,\beta}$ and $M_{\gamma,\beta}$ is of order $\Delta^{2}/g^{3}$
(see Ref.  \onlinecite{AlhassidG01} for details) which, in general, also leads
to a correction of order $\Delta/\sqrt{g}$ to the spacing. For a diffusive
QD, this term is small due to the exponential decay of $k(\br_1\smmi\br_2)$,
but it might be important in cases with strong semiclassical effects.} we get
$\text{var}(s_{\text{diag}})\simeq 2 g\, \text{var}(M_{\alpha,\beta})\propto\Delta^2
\ln{g}/g$.\cite{UllmoB01} The contribution of these terms to the spacing 
fluctuations is, then, of the same order as the scrambling, namely
$\Delta/\sqrt{g}$. A similar result can be obtained for all the different spin
transitions except for $0\!\rightarrow\!\frac{1}{2}\!\rightarrow\!0$. In
that particular case $s_{\text{diag}}\smeq M_{\frac{N}{2},\frac{N}{2}}$ and therefore
$\text{var}(s_{\text{diag}})\!\propto\!\Delta^{2}/g^{2}$. This is an exact result
not related to the zero-range approximation for $\Vtf$.  For this reason, the
effect of the fluctuation of the diagonal terms on the $\delta$-function of the
PSD is weak, though noticeable.

Diagonal terms produce fluctuations of the energy difference between the singlet
and triplet states, $E_{S=0}^{}\smmi E_{S=1}^{}$. It is straightforward
to verify that $\text{var}(E_{S=0}^{}\smmi E_{S=1}^{})\!\simeq\! 2 g\,
\text{var}(M_{\alpha,\beta})$.  Since the exchange contribution to this energy
difference is $2J_S$, we can think of this as an effective fluctuation in
$J_S$. These fluctuations might lead to a change of the ground state in the cases
where the triplet and singlet states are almost degenerate.

It is worth commenting that there is a correction of order $1/g$ to the mean 
value of the diagonal matrix elements. In the zero-range limit it is given by
\be
\langle H_{\alpha,\beta,\gamma,\delta}^{}\rangle^{(1/g)}\smeq 
		c_{1}\frac{\Delta}{2g}(\delta_{\alpha,\delta}\delta_{\beta,\gamma}\smpl 
\delta_{\alpha,\gamma}\delta_{\beta,\delta})
\ee
with $c_{1}\smeq gA^{\smmi1}\int\text{d}\br
[ \Pi_{\text{B}}(\br,\br)\smmi k ]\!\propto\!\ln{g}$.\cite{BlanterMM01} This
correction can be included in the definition of $E_C$ and $J_S$ since it has
the same structure as Eq. (\ref{mean}).  For a ballistic disc, this introduces
a correction of $\smmi0.008\Delta$ in the latter.

Before closing this subsection we should point out that $\langle M_{\alpha,\beta}
\;\mathcal{X}_{\alpha,\alpha} \rangle\smeq\mathcal{O}(1/g^2)$. In principle this
also leads to a $\Delta/\sqrt{g}$ correction to the spacing. Rough estimates
suggest that it is numerically small, however, and so we neglect it.

%%%%%%%%%%%%%%%%%%%%%%%%%%%%%%%%%%%%%%%%%%%%%%%%%%%%%%%%%%%%%%%%%%%%%%%%%
\subsection{``Gate'' effect}
%%%%%%%%%%%%%%%%%%%%%%%%%%%%%%%%%%%%%%%%%%%%%%%%%%%%%%%%%%%%%%%%%%%%%%%%%
The two contributions we have discussed so far are usually regarded as intrinsic,
in the sense that their origin is the \textit{e-e} interaction of the particles
in the QD itself.  On the other hand, the effect of the gate voltage is usually
associated with the distortion of the shape of the QD when the gate voltage is
swept.  \cite{VallejosLM98} Consequently, it appears that it  could be independently
reduced by using, for instance, QDs defined lithographically\cite{LuscherHEWB01}
or a uniform gate as opposed to a finger-shaped one. This is not true however:
the gate effect is \textit{always} as big as the scrambling effect.

The correction to the confinement potential due to a change in the gate voltage 
is\cite{AleinerBG02}
\be
\delta 
U(\br)\smeq-2E_{C}\delta\ag\smmi\Delta\left[V(\br)\delta\ag\smpl\sum_{i}{V^{(i)}(\br)\delta\ag_{i}}\right]
\label{dU}
\ee
with $\delta\ag\smeq\sum_{i}{\delta\ag^{i}}$, $\delta\ag^{i}\smeq C_{g}^{i}\,
\delta V_{g}^{i}/e$ and $\delta V_{g}^{i}$ is the change of the electrostatic
potential of the $i$-th gate. The potentials $V(\br)$ and $V^{i}(\br)$ are
related to the solution of the electrostatic problem associated with the set
of conductors (QD+gates)---see next section and Ref. \onlinecite{AleinerBG02}
for more details. The former is the \textit{same} potential that appears in
Eq. (\ref{scpotential}) and causes the scrambling effect. It is clear then that
a change in $\delta\ag$ produces the same effect as a change of $N$.  This is reasonable 
since these two effects are opposite faces of the same electrostatic problem: a
change of the electrostatic potential of the QD \textit{produces} a non-uniform
distribution of charge, which creates $V(\br)\Delta$; viceversa, an extra charge
must be distributed in the same way so that the potential of the QD is uniform.
Eq. (\ref{dU}) leads to the following correction to the Hamiltonian\cite{AleinerBG02}
\be
\Ha_{\text{gate}}^{}\!=-\sum_{\alpha,\beta,\sigma}\,{c^{\dagger}_{\alpha,\sigma}c^{}_{\beta,\sigma}
\,\left[\delta\ag\mathcal{X}^{}_{\alpha,\beta}\smpl\sum_{i}{\delta\ag_{i}\,\mathcal{X}^{i}_{\alpha,\beta}}\right]}
\ee
where $\mathcal{X}^{i}_{\alpha,\beta}$ is defined in terms of $V^{i}(\br)$---it is
then clear that $\text{var}(\mathcal{X}^{i}_{\alpha,\beta})\smeq b_{ii}\Delta^2/g$,
with $b_{ii}$ a geometry dependent numerical coefficient.

Notice that the change in the shape of the QD can be associated with the last term in
Eq. (\ref{dU}). For instance, if the potentials of two plunger-gates are swept
in a way such that $\delta\ag\smeq\delta\ag_{1}\smpl\delta\ag_{2}\smeq0$ then
the change of the confinement potential originates only from that term, and the
effect of a change in the shape of the QD can be isolated.\cite{PatelSMGASDH98}
Note, however, that this procedure actually tests only the difference between
$V^{(1)}(\br)$ and $V^{(2)}(\br)$, which could be smaller that each one of them
if the two gates are in similar positions with respect to the QD.

%%%%%%%%%%%%%%%%%%%%%%%%%%%%%%%%%%%%%%%%%%%%%%%%%%%%%%%%%%%%%%%%%%%%%%%%%
\section{Off-diagonal matrix elements.\label{sectionoff}}
%%%%%%%%%%%%%%%%%%%%%%%%%%%%%%%%%%%%%%%%%%%%%%%%%%%%%%%%%%%%%%%%%%%%%%%%%
So far we have considered only the contribution of the diagonal terms of $\Hint$.
We found that although the fluctuation of each individual matrix element is
of the order of $\Delta/g$, the total contribution to the peak spacing is of
order $\Delta/\sqrt{g}$. This is a consequence of the addition of the $\sim\!g$
different matrix elements that contribute to the spacing. Since in principle
there are many more off-diagonal terms, one might wonder if their
 contribution can also add up and result in a significant one that should also be
included.\cite{JacquodS00,JacquodS01} We show now that this is \textit{not}
the case.

The first correction to the GS energy due to off-diagonal terms appears in 
second order perturbation theory (the first order 
contribution is zero by definition)
\be
E^{(2)}_{S}\smeq\sum_{j}{\frac{\left| \langle\Psi^N_j 
|\Hint^{\text{off}}|\Psi_{S}^{N}\rangle\right|^2}{E^{(0)}_{S}-E^{(0)}_{j}}}
\label{sopt}
\ee
where $|\Psi_{S}^{N}\rangle$ is the GS of the system (described by
$\Ha_{\text{CEI}}$) with $N$ electrons and spin $S$, and $\{|\Psi_{j}^{N}\rangle\}$
are the excited states.

Following Ref. \onlinecite{JacquodS01}, we rewrite the off-diagonal part of the 
Hamiltonian (\ref{intH}) as follows,
\begin{widetext}
\bea
\nonumber
\Hint^{\text{off}}&\smeq&{\sum_{\beta\!\geq\!\alpha,\gamma\!\geq\!\delta}}^{\!\!\!\!\!'}
     {u_{\alpha,\beta,\gamma,\delta}
                     \left\{ 
c^{\dagger}_{\delta,\up}c^{\dagger}_{\gamma,\up}c^{}_{\beta,\up}c^{}_{\alpha,\up}\smpl
c^{\dagger}_{\delta,\dn}c^{\dagger}_{\gamma,\dn}c^{}_{\beta,\dn}c^{}_{\alpha,\dn}\smpl
\frac{1}{2}\;(c^{\dagger}_{\delta,\up}c^{\dagger}_{\gamma,\dn}\smpl 
c^{\dagger}_{\delta,\dn}c^{\dagger}_{\gamma,\up})
                     (c^{}_{\beta,\up}c^{}_{\alpha,\dn}\smpl 
c^{}_{\beta,\dn}c^{}_{\alpha,\up})
                     \right\}}\\
&&\qquad\qquad\smpl\frac{1}{2}\;a_{\alpha,\beta,\gamma,\delta}\;
                     (c^{\dagger}_{\delta,\up}c^{\dagger}_{\gamma,\dn}\smmi 
c^{\dagger}_{\delta,\dn}c^{\dagger}_{\gamma,\up})
                     (c^{}_{\beta,\dn}c^{}_{\alpha,\up}\smmi 
c^{}_{\beta,\up}c^{}_{\alpha,\dn})\left(1-\frac{\delta_{\alpha,\beta}}{2}\right)
                     \left(1-\frac{\delta_{\gamma,\delta}}{2}\right)
\label{off}
\eea
\end{widetext}
with 
\bea
\nonumber
u_{\alpha,\beta\gamma,\delta}&\smeq& H^{}_{\alpha,\beta,\gamma,\delta}\smmi 
H^{}_{\alpha,\beta,\delta,\gamma}\\
a_{\alpha,\beta\gamma,\delta}&\smeq& H^{}_{\alpha,\beta,\gamma,\delta}\smpl 
H^{}_{\alpha,\beta,\delta,\gamma} \,.
\eea
The sum in Eq. (\ref{off}) runs over all configurations in which the indices
of $c^{\dagger}$ and $c$ are not fully paired (that is, the configurations not
included in a Hartree-Fock treatment). From this form of the Hamiltonian, it is
easy to see that $\Hint^{\text{off}}$ conserves the total spin.\cite{JacquodS01}
The first term produces only triplet-transitions while the second only
singlet-transitions. Because of that, the states coupled by $\Hint^{\text{off}}$
have the same spin. Then, the energy denominator that appears in Eq. (\ref{sopt})
involves differences between single-electron energies. The spin rotational
invariance of $\Hint^{\text{off}}$ also implies that the second order correction to
the energy is the same for all the states in a given spin-multiplet  (\text{i.e}
the states with different $S_z$). Therefore, we can use the one with the maximum
value of $S_z$, the simplest state, throughout our calculations.

The difficulty in calculating $E^{(2)}_{S}$ lies in recognizing which terms have
to be added coherently, that is to say, which terms lead to the same final state
$|\Psi_{j}^{N}\rangle$. This complication arises because of: (i) the indices in
Eq. (\ref{off}) might be partially paired, so terms within each of the two main
terms do not necessarily produce orthogonal states, or (ii) non-trivial states,
like the  $S\smeq1$ state, can lead to the same final state under the action
of any of the two main terms in  Eq. (\ref{off}) for particular values of the
indices. In order to avoid the first problem, we explicitly take into account
all the possible pairing of the indices and rewrite Eq. (\ref{off}) as
\be
\Hint^{\text{off}}\smeq\Ha_{\text{A}}\smpl\Ha_{\text{B}}\smpl\Ha_{\text{C}}
\ee
with
\bea
\nonumber
\Ha_{\text{A}}&\smeq&\sum_{\alpha,\beta,\gamma}\sum_{\sigma}\; \big[
          u^{}_{\alpha,\beta,\gamma,\alpha}\; 
\hat{n}^{}_{\alpha,\sigma} \;
                               c^{\dagger}_{\gamma,\sigma} c^{}_{\beta,\sigma}\\
\nonumber
                      & & \!+
\frac{1}{2}\;
(u^{}_{\alpha,\beta,\gamma,\alpha}\smpl
a^{}_{\alpha,\beta,\gamma,\alpha})
\;\hat{n}^{}_{\alpha,\sigma}\; 
c^{\dagger}_{\gamma,\bar{\sigma}} c^{}_{\beta,\bar{\sigma}}\\
\nonumber
                    && \!+\frac{1}{2}\; ( 
u^{}_{\alpha,\beta,\gamma,\alpha}
\smmi a^{}_{\alpha,\beta,\gamma,\alpha})\;
c^{\dagger}_{\alpha,\sigma} 
c^{}_{\alpha,\bar{\sigma}} c^{\dagger}_{\gamma,\bar{\sigma}} 
                                 c^{}_{\beta,\sigma} \big] \\
\nonumber
                      & & 
+ \sum_{\alpha,\beta}{\sum_{\sigma}{\frac{1}{2}\;
(a^{}_{\alpha,\beta,\beta,\beta}\smpl a^{}_{\beta,\beta,\beta,\alpha})\;
                                  \hat{n}^{}_{\beta,\sigma}\; 
c^{\dagger}_{\beta,\bar{\sigma}} c^{}_{\alpha,\bar{\sigma}}
                               }}\; ,\\
\eea
\bea
\nonumber
\Ha_{\text{B}}&\smeq&\sum_{\alpha,\gamma}{\frac{1}{2}\; 
a^{}_{\alpha,\alpha,\gamma,\gamma}\; 
c^{\dagger}_{\gamma,\up}c^{\dagger}_{\gamma,\dn}c^{}_{\alpha,\dn}c^{}_{\alpha,\up}}\\
\nonumber
                      &   &        \smpl 
\sum_{\alpha,\gamma>\delta}{\frac{1}{2}\; a^{}_{\alpha,\alpha,\gamma,\delta}\; 
(c^{\dagger}_{\delta,\up}c^{\dagger}_{\gamma,\dn}\smmi 
c^{\dagger}_{\delta,\dn}c^{\dagger}_{\gamma,\up})
                                      c^{}_{\alpha,\dn}c^{}_{\alpha,\up}} \\
\nonumber
                      &  &         \smpl\sum_{\gamma,\beta>\alpha}{\frac{1}{2}\; 
a^{}_{\alpha,\beta,\gamma,\gamma}\; 
c^{\dagger}_{\gamma,\up}c^{\dagger}_{\gamma,\dn}
                                     (c^{}_{\beta,\dn}c^{}_{\alpha,\up}\smmi 
c^{}_{\beta,\up}c^{}_{\alpha,\dn})
                                     }\\
\eea
and $\Ha_{\text{C}}$ as in Eq. (\ref{off}) but with all the indices
being different. Here, none of the remaining indices are paired and
$\bar{\sigma}\smeq\smmi\sigma$. Notice that $\Ha_{\text{A}}$ contains contributions
from both terms in Eq. (\ref{off}).

In general, the eigenstates of $\Ha_{\text{CEI}}$ are a superposition of
Hartree-Fock states. However, because $\vec{S}^{2}$ and the single particle
Hamiltonian commute, they can be classified by their occupation numbers
$\{n_{\alpha}\}$. For each configuration with $N_s$ singly occupied levels,
there are $2^{N_s}$ states which have different values of $S$ and $S_z$.  It is
worth mentioning that for $N_s\!\ge\!3$, the values of $S$ and $S_z$ are not
enough to specify a given eigenstate--for instance, for $N_s\!\smeq\!3$, there
are two orthogonal sets of states with $S\smeq\frac{1}{2}$.  The determination
of the precise form of the spin eigenstates in terms of the HF states is not a
trivial task.\cite{Pauncz} Nevertheless, for our purpose, it is sufficient to
notice that in order to add coherently two terms must lead to final states with
the same occupation numbers.  With this in mind, it is straightforward to check
that $\Ha_{\text{A}}$, $\Ha_{\text{B}}$ and $\Ha_{\text{C}}$ add incoherently. In
analyzing each of them however, we must proceed in a case by case basis.

At present, we are not aware of any simple method for evaluating Eq. (\ref{sopt})
for an arbitrary state $|\Psi_{S}^{N}\rangle$. However, it is sufficient for our
purpose to calculate the correction for the cases $S\smeq0$, $\frac{1}{2}$ and
${1}$ since those are the most probable values of the spin for typical values of
the \textit{e-e} interaction.  As an example, let us discuss the $S\smeq0$ case in
detail. In this case, $|\Psi_{S=0}^{N}\rangle$ has only doubly occupied levels
(up to $\Ef$).  It is easy to see that the last two terms of $\Ha_{\text{A}}$
give zero when applied on $|\Psi_{S=0}^{N}\rangle$. The remaining terms of
$\Ha_{\text{A}}$ give,
\bea
\nonumber
\Ha_{\text{A}}|\Psi_{S=0}^{N}\rangle&\smeq&\sum_{\beta,\gamma}\left(\sum_{\alpha}\frac{1}{\sqrt{2}}(3u^{}_{\alpha,\beta,\gamma,\alpha}\smpl 
    a^{}_{\alpha,\beta,\gamma,\alpha})\right)\\
       &&\times\frac{c^{\dagger}_{\gamma,\up} c^{}_{\beta,\up}\smpl 
c^{\dagger}_{\gamma,\dn} c^{}_{\beta,\dn}}{\sqrt{2}}\, |\Psi_{S=0}^{N}\rangle
\eea
Note that the sum over $\alpha$ does not affect the final states---this is an 
example of terms that add coherently---while each pair $\{\beta,\gamma\}$ 
leads to a different final state. The factor $\sqrt{2}$ was introduced to keep 
the final state properly normalized. 
Similarly, it is easy to check that each of the terms in $\Ha_{\text{B}}$ and 
$\Ha_{\text{C}}$ lead to orthogonal states. Adding up all the contributions, we 
finally get the 
following expression, 
\bea
\nonumber
E^{(2)}_{S=0}\smeq\sum_{\beta,\gamma}{\frac{\left| \sum_{\alpha\neq\beta} 3 
u_{\alpha,\beta,\gamma,\alpha}\smpl a_{\alpha,\beta,\gamma,\alpha}\right|^{2}}
				{2(\ve_{\beta}\smmi\ve_{\gamma})}}\\
\nonumber
      \smpl \sum_{\beta\geq\alpha,\gamma\geq\delta}{\frac{3\left| 
u_{\alpha,\beta,\gamma,\delta}\right|^{2}
                \smpl\left| 
a_{\alpha,\beta,\gamma,\delta}\right|^{2}}{\ve_{\alpha}\smpl\ve_{\beta}\smmi\ve_{\gamma}\smmi\ve_{\delta}}
    (1\smmi\frac{\delta_{\alpha,\beta}}{2})(1\smmi\frac{\delta_{\gamma,\delta}}{2})}\\
\eea
where $\alpha$ and $\beta$ ($\gamma$ and $\delta$) refer to occupied
(empty) levels. Notice that the first (coherent) term is absent in
Ref. \onlinecite{JacquodS01}.  Following a similar procedure, we have obtained
expressions for $S\smeq\frac{1}{2}$ (see Appendix) and $S\smeq1$ which are too
cumbersome to be presented here. In those cases, however, the presence of single
occupied levels introduces some complications, and special care must be taken with
the terms in which any of the indices correspond to one of these levels.  It is
also important to properly take into account the symmetry properties of the matrix
elements, $H_{\alpha,\beta,\gamma,\delta}\smeq H_{\beta,\alpha,\delta,\gamma}$,
as it affects the variance of $u_{\alpha,\beta,\gamma,\delta}$ and
$a_{\alpha,\beta,\gamma,\delta}$.  Note than in the particular case of the
zero-range interaction limit, $u_{\alpha,\beta,\gamma,\delta}\!\equiv\!0$ and
$a_{\alpha,\beta,\gamma,\delta}\smeq2H_{\alpha,\beta,\gamma,\delta}$

We calculate $E_{S}^{(2)}$ numerically in the zero-range limit.  We first
evaluate the correction to the spin gap, $\Delta_{S}\smeq E^{(2)}_{S=0}\smmi
E^{(2)}_{S=1}$.  We find that both $\langle\Delta_S\rangle$ and 
$\text{rms}(\Delta_{S})$ are of order $\Delta/g$
which, being a higher order correction, can be neglected. Nevertheless, we checked
that for $N\smeq500$, $\text{rms}(\Delta_{S})$ is $\sim\!4$ times smaller than
the correction introduced by the diagonal terms. Therefore, we ignored any
effect of the off-diagonal terms in the occupation of the states. This allows
us to simply evaluate the correction to the spacing as $s_{\text{off}}\smeq
E^{(2)}_{N\smpl1,S''}\smpl E^{(2)}_{N\smmi1,S'}\smmi2E^{(2)}_{N,S}$.
We find that $\text{var}({s_{\text{off}}})\!\propto\!\Delta^2/g^2$ and so
can also be neglected. An explicit evaluation for $N\smeq500$ gives
$\text{rms}({s_{\text{off}}})\smeq0.016\Delta$ which is much smaller than the
fluctuation introduced by the diagonal terms. The latter result corresponds
to a generic transition, while for the special case 
$0\!\rightarrow\!\frac{1}{2}\!\rightarrow\!0$
we get $\text{rms}({s_{\text{off}}})\smeq0.007\Delta$.

%
%%%%%%%%%%%%%%%%%%%%%%%%%%%%%%%%%%%%%%%%%%%%%%%%%%%%%%%%%%%%%%%%%%%%%%%%%
\section{Magnitude of Scrambling Effect\label{evalvarchi}}
%%%%%%%%%%%%%%%%%%%%%%%%%%%%%%%%%%%%%%%%%%%%%%%%%%%%%%%%%%%%%%%%%%%%%%%%%
%
As we mentioned in the previous section, the origin of the potential $V(\br)$ in 
Eq. (\ref{scpotential}) is the screening charge
$-e/\kappa$. Although it has been known for a while\cite{BlanterMM97} that this 
leads to a correction of order $\Delta/\sqrt{g}$ to the Hamiltonian, 
$\text{var}(\mathcal{X}_{\alpha,\beta})\smeq b_{00}\Delta^2/g$, a 
realistic estimate of the magnitude of $b_{00}$ is 
still lacking. This is particularly important since the effect of the scrambling 
on the PSD goes in the right direction, i.e. it can 
lead to Gaussian-like distribution if it is strong enough. Here, we show that 
this has been overestimated in the literature\cite{AlhassidM99,VallejosLM99} and 
that 
scrambling is not able by itself to explained the experimental results.
  
The evaluation of $b_{00}$ for actual geometries is quite 
difficult. The reason is that it involves finding the solution 
of the electrostatic field for a set of conductors in a particular 
geometry.\cite{AleinerBG02} Following Ref. \onlinecite{AleinerBG02} 
we write
\be
V(\br)\smeq\frac{A\kappa}{8\pi C}\int \text{d}\br_2 \tensor{\partial}_z 
\tensor{\partial}_{z_2} \mathcal{D}(\br,\br_2)
\label{defscpotential}
\ee
where $\tensor{\partial}_z\smeq\partial_{z_{\text{-}}}-\partial_{z_{\text{+}}}$ 
($z$ is the axis perpendicular to the dot), 
$\mathcal{D}(\br_1,\br_2)$ is the Green function of the electrostatic problem 
outside the QD \textit{including} the gates,
\be
\nabla^2_{\br}\mathcal{D}(\br,\br_2)\smeq-\delta(\br\smmi\br_2),\qquad\mathcal{D}(\br,\br_2)|_{\br\in\mathcal{S}}\smeq0
\ee
with $\mathcal{S}$ the conducting surfaces and 
\be
C\smeq\left|\frac{\kappa}{4\pi }\int \text{d}\br_1\text{d}\br_2 
\tensor{\partial}_{z_1} \tensor{\partial}_{z_2} \mathcal{D}(\br_1,\br_2)\right|
\label{C}
\ee
is the total capacitance of the QD. Eq. (\ref{defscpotential}) has a very clear 
physical interpretation if we notice that 
\be
 \phi(\br)\smeq\frac{-e}{C}\int \text{d}\br_2 \tensor{\partial}_{z_2} 
\mathcal{D}(\br,\br_2)
 \label{potential}
\ee
is the electrostatic potential outside the QD with 
$\phi(\br)|_{\br\in\text{QD}}\smeq -e/C$ and $\phi(\br)\smeq0$ over the gate 
electrodes. Then, 
\be
V(\br)\smeq-\frac{A\kappa}{2 e}\sigma(\br)	
\label{scrampotential}
\ee
with 
\be
\sigma(\br)\smeq\frac{1}{4\pi}\tensor{\partial}_z \phi(\br)	
\label{charge}
\ee
the surface charge density in the QD associated with electrostatic potential 
$\phi(\br)$. Using Eq. (\ref{C}), it is straightforward to show that 
$Q\smeq \int \text{d}\br \sigma(\br)\smeq -e/\kappa$; note that this implies
$\bar{V}\smeq\frac{1}{2}$. Introducing Eq. (\ref{scrampotential}) in Eq. 
(\ref{varchi}) and using the fact that 
$k(\br_1,\br_2)\approx1/\pi\kf|\br_1\smmi\br_2|$ we get 
\bea
\nonumber
\text{var}(\mathcal{X}_{\alpha,\beta})&\smeq&
   \frac{\Delta^2}{4\pi\kf 
\sqrt{A}}\Bigg(\frac{1}{A^{\frac{3}{2}}}\int\frac{\text{d}\br_1\text{d}\br_2}{|\br_1-\br_2|}\\
  &\smpl& 
\frac{\sqrt{A}}{Q^2}\int\text{d}\br_1\left[\sigma(\br_1)\smmi\frac{2Q}{A}\right]\phi'(\br_1)
\Bigg)
\label{varchipotential}
\eea
where
\be
\phi'(\br_1)\smeq\int\text{d}\br_2 \frac{\sigma(\br_2)}{|\br_1-\br_2|}
\ee
is the potential due to the surface charge in the QD. 
Let us now consider the different cases.

\subsection{Isolated dot}
In this case, the only charge in the system is $\sigma(\br)$ and the 
electrostatic potential $\phi'(\br)$ is constant 
over the QD surface, $\phi'(\br)\smeq\phi(\br)\smeq-e/C$.  This allow us to 
readily obtain\cite{UllmoB01} 
\be
\text{var}(\mathcal{X}_{\alpha,\beta})\smeq
        \frac{\Delta^2}{4\pi\kf 
\sqrt{A}}\left(\alpha\smmi\frac{\sqrt{A}\kappa}{C}\right),
\label{varchiisolated}
\ee
with $\alpha\smeq 
A^{\smmi\frac{3}{2}}\int\text{d}\br_1\text{d}\br_2\,|\br_1-\br_2|^{\smmi1}$. 
Notice that no particular geometry has been assumed so far. The first term in the 
parenthesis can be calculated numerically for 
arbitrary geometries. The second, however, requires the calculation of the 
capacitance. In the case of an ellipsoidal QD,\cite{Landau_Vol8}
\be
\frac{\sqrt{A}\kappa}{C}\smeq\sqrt{\frac{b}{a}}\frac{\sqrt{\pi}}{x}F(\arcsin{x},\frac{1}{x^2})\;,\qquad 
x\smeq\sqrt{1\smmi\left(\frac{b}{a}\right)^2}
\ee
where $a$ ($b$) is the length of the long (short) axis and
$F(\phi,m)$ is the elliptic integral of the first kind.  For
$a/b\simeq1-3$ we get $b_{00}\simeq0.002$. It is clear then that
$\text{rms}(\mathcal{X}_{\alpha,\beta})\!\simeq\!0.04\Delta/\sqrt{g}$
is smaller than usually assumed.  In fact, for $N\smeq500$, this value
corresponds to $\delta x\smeq \sqrt{b_{00}/g}\smeq0.018 $ in the parametric
approach to the scrambling, about a factor of $10$ less than taken in Refs.
\onlinecite{VallejosLM98} and \onlinecite{AlhassidM99}.

\subsection{Dot with gates}
Since experiments are certainly done in the presence of gate electrodes,  a careful
calculation should take them into account.  Looking at Eq. (\ref{varchiisolated}),
it is tempting to simply replace $C$ by its experimental value. However, the
above calculation is only valid for an isolated QD, as we explicitly assumed that
$\sigma(\br)$ was the only charge in the systems. For a real QD, the induced charge
on the gates has to be considered. Then, $\phi(\br)\smeq\phi'(\br)\smpl\phi''(\br)$
where $\phi''(\br)$ is the potential created by the induced charge on the gates.
Defining $\bar{\phi''}\smeq A^{\smmi1}\int\text{d}\br \phi''(\br)$ and $\beta\smeq
(Q\bar{\phi''})^{\smmi1}\int\text{d}\br \sigma(\br)\,\phi''(\br)$ we find
\be
\text{var}(\mathcal{X}_{\alpha,\beta})\smeq  
\frac{\Delta^2}{4\pi\kf 
\sqrt{A}}\left(\alpha\smmi\frac{\sqrt{A}\kappa}{C}\left[1\smpl(2\smmi\beta)\frac{\bar{\phi''}}{e/C}\right]\right)
\label{varchifinal}
\ee
Notice that the value of the ratio $\sqrt{A}\kappa/C$ can now be
obtained from the experimental data since $C$ is the capacitance for
the actual geometry.  

Since $\beta\!\sim\!1$ and $\bar{\phi''}\!>\!0$
(the sign of the induced charge is the opposite of $Q$), it is evident
that using the isolated dot result [Eq.  (\ref{varchiisolated})] gives an
upper limit to $\text{var}(\mathcal{X}_{\alpha,\beta})$ when evaluated
with the experimental parameters.  An estimate of $\bar{\phi''}$ is
obtained as follows. Let $Q_{i}$ be the charge of the $i$-th gate, then
$\bar{\phi''}\!\sim\!\sum_{i}{Q_{i}/d_{i}\kappa}$, with $d_{i}$ the distance
between the centers of charge of the QD and the $i$-th gate.  Since $Q_{i}\smeq
\smmi C^{i}_g (-e/C)$, where $C_g^{i}$ is the dot-$i$-th gate capacitance, it
turns out that $\bar{\phi''}/(e/C)\!\sim\!\sum_{i} C_g^{i}/d_i\kappa$.  Then,
\be
\frac{\sqrt{A}\kappa}{C} \frac{\bar{\phi''}}{e/C}\!\sim\!0.5 \frac{\sqrt{A}}{d}
\ee
where we used that typically\cite{WaughBMW95} $\sum_{i} C_g^{i}/C\!\sim0.5$ and
denote by $d$ the average distance between the center of the dot and the gates. For
the data of Ref. \onlinecite{PatelCSHMDHCG98} we estimate $b_{00}\!\sim\!0.005$
with an upper limit of $0.01$; thus, even with gate effects included,
our estimate is smaller than values used previously.\cite{VallejosLM98,AlhassidM99}

%%%%%%%%%%%%%%%%%%%%%%%%%%%%%%%%%%%%%%%%%%%%%%%%%%%%%%%%%%%%%%%%%%%%%%%%%
\section{Ground state peak spacing distribution \label{results}}
%%%%%%%%%%%%%%%%%%%%%%%%%%%%%%%%%%%%%%%%%%%%%%%%%%%%%%%%%%%%%%%%%%%%%%%%%
We now use numerics for the evaluation of the PSD. At $T\smeq0$ (temperature 
effects will be discussed in the next section) the peak spacing is given by
\be
s_{N}\smeq [\Egs^{N\smpl1}(\ag')\smmi\Egs^{N}(\ag')]\smmi 
[\Egs^{N}(\ag)\smmi\Egs^{N\smmi1}(\ag)]
\ee
where 
$\Egs^N(\ag)$ is the GS energy of the QD with $N$ electrons \textit{excluding} 
the charging energy term and  $\ag$ is the corresponding gate voltage. 
\footnote{With this definition, $s_N$ has the classical result $2E_C$ 
subtracted. That is, $s_N\smeq0$ ($\Delta\ve$) for the odd (even) case in the 
CI model.}
Including all the leading order corrections to the CEI model, the Hamiltonian 
reads\cite{AleinerBG02} 
\begin{eqnarray}
\Ha_\text{QD}&\smeq& \Ha_\text{CEI}^{}
               \smpl\frac{1}{2}
\sum_{\alpha,\beta,\gamma,\delta}^{\text{diag}}{H_{\alpha,\beta,\gamma,\delta}^{}\;
               \; c^{\dagger}_{\delta,\sigma}c^{\dagger}_{\gamma,\sigma'}
               c^{}_{\beta,\sigma'}c^{}_{\alpha,\sigma}}
\nonumber \\
        &     & 
\smpl\sum_{\alpha,\beta,\sigma}\,{c^{\dagger}_{\alpha,\sigma}c^{}_{\beta,\sigma}
                 \,[(\hat{n}-\ag){\cal X}^{}_{\alpha,\beta}
                 \smpl\delta\ag\,{\cal X}^{1}_{\alpha,\beta}}]
\label{leading}
\end{eqnarray}
where the variance of $H_{\alpha,\beta,\gamma,\delta}^{}$ and ${\cal
X}^{j}_{\alpha,\beta}$ are given by Eq. (\ref{varM1}) and Eq. (\ref{varchifinal}),
respectively.  Their mean values are included in the definition of $E_C$ and
$J_S$, so that $\langle H_{\alpha,\beta,\gamma,\delta}^{}\rangle\smeq\langle{\cal
X}^{j}_{\alpha,\beta}\rangle\smeq0$.  The second term in Eq. (\ref{leading})
includes only the diagonal terms of the residual interaction, and $\delta\ag$
is taken with respect to some fixed state, for instance the state with $N$
electrons.  The GS energies were obtained by minimizing the energy with respect
to the occupation numbers for $N\smmi1$, $N$, $N\smpl1$, and $N\smpl2$ (and the
corresponding $\ag$, $\ag'$, and $\ag''$). We only kept two consecutive spacings
for each realization of the single-particle Hamiltonian.  It is worth mentioning
again that Eq. (\ref{leading}) is defined in a window up to $\Eth$, so that only
$g$ levels were considered.

The parameters we use are: (1) $g\smeq0.38\sqrt{N/2}$ (which corresponds to a
disc geometry);\cite{AleinerBG02,BlanterMM97} (2) the upper limit for the value
of $b_{00}\!\simeq\!0.01$,  and (3) $b_{11}\smeq b_{00}$.  Only transitions with
$|\delta S|\smeq\frac{1}{2}$ were taken into account  since in the absence of
spin-orbit interaction (or if it is small), the transitions where the change
in the GS spin is bigger than $\frac{1}{2}$, appear as ``missing'' peaks in the
conductance and are not included in the experiment.

\begin{figure}[t]
\begin{center}
 \includegraphics[width=8.3cm,clip] {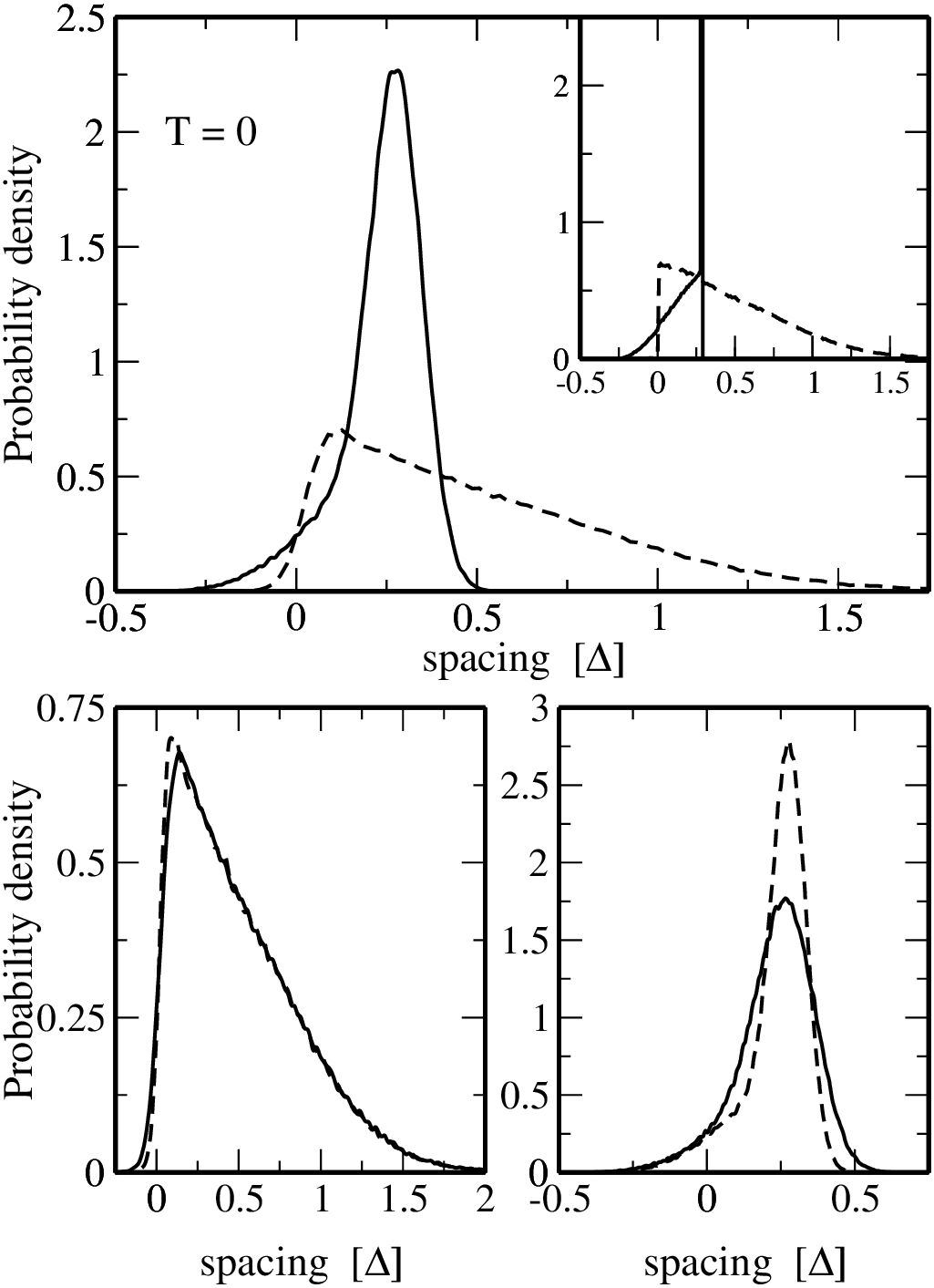}
\end{center}
\caption{Top panel: Ground state  peak spacing distribution obtained from Eq. 
(\ref{leading}) with $N\smeq500$ ($g \!\approx\! 6$) and $J_S \smeq 0.28 \Delta$ 
($r_s\smeq1$). The solid (dashed) line correspond to $N$ odd (even). The origin 
of the horizontal axis corresponds to $s_{N}\smeq J_{S}/2$. 
The CEI model result is included in the inset for comparison. Note that all the 
sharp features, 
including the $\delta$-function, are washed-out but the long tail for large 
spacing persists in the even case. Bottom panel: Comparison of the even (left)and 
odd (right) 
distributions for $N\smeq200$ (solid line) and $N\smeq1000$ (dashed line). 
Clearly, residual effects are stronger for $N$ odd.} 
\label{GSPSD}
\end{figure}
\begin{table}[b]
\caption{\label{table1}Comparison of the $1/\sqrt{g}$ corrections for different 
number of electrons, $N$. We use the relation $g\simeq0.27\sqrt{N}$, which is 
valid for a disc 
geometry.\cite{BlanterMM98}
}
\begin{ruledtabular}
\begin{tabular}{ccccc}
$N$ 
&$g$&$\text{rms}(\mathcal{X}_{\alpha,\beta})$&$\text{rms}(M_{\alpha,\beta})$&$\text{rms}(s_N)$ 
\\ 
\hline
 $200$&$3.8$ &$0.051\Delta$&$0.031\Delta$ &$0.313\Delta$ \\
\hline
 $500$&$6.1$ &$0.041\Delta$&$0.020\Delta$ &$0.308\Delta$ \\ 
\hline
 $1000$&$8.6$&$0.034\Delta$ &$0.015\Delta$ &$0.305\Delta$ \\ 
\end{tabular}
\end{ruledtabular}
\end{table}
Numerical results for $N\smeq200$, $500$, and $1000$ are shown in Fig. 
\ref{GSPSD}. 
The horizontal axis corresponds to $s_{N}\smmi J_{S}/2$, so that the origin 
agrees with the Hartree-Fock result of Ref. \onlinecite{UllmoB01}. 
Corrections to the CEI model clearly smear-out all the sharp features of the PSD. 
The former $\delta$-function (see inset in Fig. \ref{GSPSD}) is now a finite peak 
but 
still constitutes the dominant feature of the PSD. As expected, the additional 
corrections increase the 
r.m.s. of the spacing with respect to its value in the CEI model ($0.28\Delta$) 
(see Table \ref{table1}). 
Notice that $1/\sqrt{g}$ corrections mainly affect the odd distribution---the 
even distribution remains essentially unaltered. This is evident in the 
comparison between the 
$N\smeq200$ and $N\smeq1000$ cases (see bottom panel in Fig \ref{GSPSD}).  
Therefore, both the scrambling and the gate effect are the dominant effects as 
they are the ones that most affect the $\delta$-function---the effect of the 
diagonal terms on the $\delta$-function is higher order in $1/g$ (see discussion 
in Section \ref{diag}). 

It is important to emphasize that besides the smearing caused by the additional 
corrections, there is still a noticeable even/odd effect. This theory  then 
\textit{does} indeed 
predict such an effect at \textit{low} $T$. To what extent this is valid at 
higher $T$ is discussed next.

%%%%%%%%%%%%%%%%%%%%%%%%%%%%%%%%%%%%%%%%%%%%%%%%%%%%%%%%%%%%%%%%%%%%%%%%%
\section{Temperature effects \label{temp}}
%%%%%%%%%%%%%%%%%%%%%%%%%%%%%%%%%%%%%%%%%%%%%%%%%%%%%%%%%%%%%%%%%%%%%%%%%
So far we have ignored thermal excitations and calculated the PSD in terms of GS 
energies (Fig.\ref{GSPSD}). This will remain a good approximation so long as 
$\kt\!\ll\!\delta$, where $\delta$ is the energy difference between the  GS and 
the first excited state. In that case, the contribution from the excited states 
can be ignored. This has been an implicit assumption in most previous work (note 
however Refs. \onlinecite{PatelSMGASDH98} and \onlinecite{AlhassidM99}). 
We will show now that in the CEI model the condition for achieving 
$\kt\!\ll\!\delta$ is much more restrictive than in the CI model and that in fact 
temperature effects 
are crucial for understanding the experimental data.\cite{UsajB01_RC}  

There are two simple reasons to expect stronger temperature effects in the 
presence of exchange. First, there is a change in the occupation of the excited 
states. 
For example, assuming $N$ even, we have $\delta\smeq|\Delta\ve\smmi2J_S|$, with 
$\Delta\ve$ the single-particle energy spacing between the two top levels. 
A significant occupation of the first excited level occurs when $\kt\!\sim\!\delta$. 
Since we are assuming $\kt\!\ll\!\Delta$, this implies $\delta\!\ll\!\Delta$.
While in the CI model  the probability for that to occur is small due to level 
repulsion, this is not 
so in the presence of exchange (i.e. $\Delta\ve\!\sim\!2J_S$ is much more likely 
than $\Delta\ve\!\sim\!0$).\cite{UsajB01_RC}
Second, at finite temperature the peak position involves the change in {\em free 
energy} of the QD upon adding a particle. 
Then, as we show below, the entropy 
contribution\cite{GlazmanM88,Beenakker91,Akera99a} leads to a shift of the peak 
position that depends on the spin 
transition in the QD.
%
%%%%%%%%%%%%%%%%%%%%%%%%%%%%%%%%%%%%%%%%%%%%%%%%%%%%%%%%%%%%%%%%%%%%%%%%%%
\subsection{General approach}
%%%%%%%%%%%%%%%%%%%%%%%%%%%%%%%%%%%%%%%%%%%%%%%%%%%%%%%%%%%%%%%%%%%%%%%%%%
%
We now proceed with a detailed calculation. Let us consider the regime $\Gamma\ll 
\kt,\Delta\ll E_C$, where 
$\Gamma$ is the total width of a level in the QD. Near the CB peak corresponding 
to the 
$N\smmi1\!\rightarrow\!N$ transition, the linear conductance is given by
\cite{Beenakker91,MeirW92} 
\be
G(\ag)\smeq\frac{e^2}{\hbar \kt}\, P_\text{eq}^{N}\,
                \sum_{\alpha}{\frac{\Gamma_{\alpha}^L\Gamma_{\alpha}^R}
                 {\Gamma_{\alpha}^L\smpl\Gamma_{\alpha}^R}\,w_{\alpha}}
\label{G}
\ee
with $\Gamma^{L(R)}_{\alpha}$ the partial width of the single-particle level 
$\alpha$ due to tunneling to the left (right) lead and
$w_{\alpha}$ a weight factor given by
\be
w_{\alpha}\smeq\sum_{i,j,\sigma}{F_\text{eq}(j|N)
\left|\langle\Psi_{j}^{N}|c_{\alpha,\sigma}^{\dagger}|\Psi_{i}^{N\smmi1}\rangle\right|^2
              [1\smmi f(\e_{j}\smmi\e_{i})]}.
\label{weight}
\ee
Here, (1) $P_\text{eq}^{N}$ is the {\em equilibrium} probability that the QD 
contains $N$ electrons, (2)  
$\Ha_\text{QD}|\Psi_{j}^N\rangle\smeq\e_{j}|\Psi_{j}^N\rangle$ so that ``$j$'' 
labels the many-body states of the QD, 
(3) $F_\text{eq}(j|N)$ is the conditional probability that the eigenstate $j$ is 
occupied given 
that the QD contains $N$ electrons, and (4) 
$f(\e)\smeq\{1\!+\exp[(\e-\!\Ef)/\kt]\}^{-1}$. 
Since near the peak only the states with $N\smmi1$ and $N$ electrons are 
relevant, we have\cite{Beenakker91}
\bea
\nonumber
P_\text{eq}^{N}&\!\simeq\!&\frac{\exp{(-\Omega_N)}}{\exp{(-\Omega_N)}\smpl\exp{(-\Omega_{N\smmi1})}}\\
                       &\smeq&f({\cal F}_{N}\smmi{\cal F}_{N\smmi1})
\eea
with $\Omega_N$ the grand-canonical potential and ${\cal F}_N$ the canonical free 
energy of the QD.
To make the dependence on $\ag$ explicit, let us denote by $\{E_{j}^{}\}$ the 
eigenenergies of 
$\Ha_\text{QD}$ without the charging energy term and define 
$\delta\ag\smeq(N\smmi\frac{1}{2})\smmi\ag$. Then, 
$\e_{j}\smmi\e_{i}\smeq E_{j}^{N}\smmi E_{i}^{N\smmi1}\smpl 2E_C \delta\ag$ and 
\bea
\nonumber
\mathcal{F}_{N}\smmi\mathcal{F}_{N\smmi1}&\smeq&\smmi\kt\ln\left[\frac{\mathcal{Z}_{N}}{\mathcal{Z}_{N\smmi1}}\right] 
\\
\nonumber
     &\smeq&E_{j}^{N}\smmi 
E_{i}^{N\smmi1}\smpl\kt\ln\!\left[\frac{F_\text{eq}(j|N)}{F_\text{eq}(i|N\smmi1)}\right]\\
&&\smpl 2E_C \delta\ag
\label{freeE}
\eea
with  $\mathcal{Z}_{N}$ the canonical partition function. Note that Eq. 
(\ref{freeE}) is valid for any $i$ and $j$. 
The contribution of the transition $i\!\rightarrow\!j$ to the conductance reaches 
its maximum when 
$f({\cal F}_{N}\smmi{\cal F}_{N\smmi1})[1\smmi f(\e_{j}\smmi\e_{i})]$ peaks, 
namely when
\be
\Ef\smeq E_{j}^{N}\smmi E_{i}^{N\smmi1}
\smpl\frac{\kt}{2}\ln\left[\frac{F_\text{eq}(j|N)}{F_\text{eq}(i|N\smmi1)}\right]
\smpl2E_C\delta\ag.
\label{gcond} 
\ee
%
%%%%%%%%%%%%%%%%%%%%%%%%%%%%%%%%%%%%%%%%%%%%%%%%%%%%%%%%%%%%%%%%%%%%%%%%%%
\subsection{Ground state dominated transitions}
%%%%%%%%%%%%%%%%%%%%%%%%%%%%%%%%%%%%%%%%%%%%%%%%%%%%%%%%%%%%%%%%%%%%%%%%%%
%
In the particular case where the  transition between GS dominates, and taking the 
spin degeneracy into account, the CB peak 
position is given by
\be 
\Ef\smeq\Egs^{N}\smmi\Egs^{N\smmi1}\smmi\frac{\kt}{2}
\ln\left[\frac{2S_{\text{GS}}^N\smpl1}{2S_{\text{GS}}^{N\smmi1}\smpl 
1}\right]\smpl2E_C\delta\ag.
\label{cond}
\ee
We see that the peak is \textit{shifted} with respect to its position
at $T\smeq0$ by an amount depending on the change of the spin of the
QD.\cite{GlazmanM88,Beenakker91,Akera99a} Except for a factor $\frac{1}{2}$ in front
of the entropic term, Eq. (\ref{cond}) corresponds to replacing $E_{\text{GS}}^{N}$
by $\mathcal{F}_{N}$ in the usual condition for the peak position, which is what
we would naively expect at finite temperature.  Because the r.m.s. of the PSD
is $\!\sim\!0.3\Delta$ (see Fig.~1), this shift is significant \textit{even}
for $\kt\!\sim\!0.1\Delta$ and cannot be neglected.  Notice that we have not
made any assumptions about the Hamiltonian of the QD so far ---except that
close to the conductance peak it depends on the gate voltage only through
the charging term.  \footnote{This is no longer true once the gate effect is
introduced. However, taking this into account would introduce a correction of order
$(\Delta/E_C)\times \text{rms}(\mathcal{X}_{\alpha,\beta})$, which is negligible.}
While in the CI model this introduces only a constant shift between the even
and odd distributions,\cite{KaminskiG00} in the CEI model it changes the {\em
shape} of both distributions since different spin transitions contribute to each
one. Also, one should note that this entropic effect shifts the energy $\Egs^{N}$
in the same direction as the exchange interaction. Then, we should expect an effect
on the PSD similar to the one corresponding to an effective increase of $J_S$.

%
%%%%%%%%%%%%%%%%%%%%%%%%%%%%%%%%%%%%%%%%%%%%%%%%%%%%%%%%%%%%%%%%%%%%%%%%%%
\subsection{Peak conductance}
%%%%%%%%%%%%%%%%%%%%%%%%%%%%%%%%%%%%%%%%%%%%%%%%%%%%%%%%%%%%%%%%%%%%%%%%%%
%
It is important to point out that the magnitude of the on-peak conductance is 
renormalized because of the spin 
degeneracy.\cite{GlazmanM88,Beenakker91} The reason is that $F_\text{eq}(j|N)$, 
the overlap 
$\left|\langle\Psi_{j}^{N}|c_{\alpha,\sigma}^{\dagger}|\Psi_{i}^{N\smmi1}\rangle\right|^2$ 
and the value of 
$P_\text{eq}^{N}\!\times\![1\smmi f(\e_{j}\smmi\e_{i})]$ at its maximum depend on 
the particular spin transition involved. 
In the simplest case when only the GS is relevant, we get
%f({\cal F}_{N}\smmi{\cal F}_{N\smmi1})
\be
F_\text{eq}(j|N) \{P_\text{eq}^{N}[1\smmi 
f(\e_{j}\smmi\e_{i})]\}|_{\text{max}}\smeq
\frac{1}{\left(\sqrt{2S'\smpl1}\smpl\sqrt{2S\smpl1}\right)^{2}}
\ee
and 
\be
\sum_{S_{z}',k';S_{z},k;\sigma}\left|\langle\Psi_{S_z',k'}^{N}|c_{\alpha,\sigma}^{\dagger}|\Psi_{S_z,k}^{N\smmi1}\rangle\right|^2
\smeq\left\{
\begin{array}{ll}
2S'\smpl1& \text{if $n_{\alpha}\smeq0$}\\
&\\
2S\smpl1&\text{if $n_{\alpha}\smeq1$}\\
\end{array}
\right.
\label{overlap}
\ee
where $S'\,(S)$, $S_z'\,(S_z)$ and $k'\,(k)$ are the quantum numbers associated 
to the state with $N\, (N\smmi1)$ particles.
\footnote{The quantum numbers $S$ and $S_z$ are not enough to specify 
a many-body state if  the number of singly occupied levels is bigger than 2, even 
for a given set $\{n_{\alpha}\}$ of occupation numbers (see Ref. 
\onlinecite{Pauncz}).}  
At low temperature, most transitions correspond to the first case in Eq. 
(\ref{overlap}) when $S'\!>\!S$ and to the second when 
$S'\!<\!S$. Using that, we finally get  
\be
G_{\text{peak}}\smeq\lambda\frac{2e^2}{\hbar \kt}\, 
\frac{\Gamma_{\alpha}^L\Gamma_{\alpha}^R}{\Gamma_{\alpha}^L\smpl\Gamma_{\alpha}^R}.
\label{maxconduct}
\ee
with
\be
\lambda\smeq  \frac{2(S'\smpl S)\smpl 3}
                    {4\,\left(\sqrt{2S'\smpl1}\smpl\sqrt{2S\smpl1}\right)^{2}}
\ee
Then, the average conductance peak depends not only on the average coupling to the
leads but also on the probability of the transition $S\!\rightarrow\!S'$---i.e.,
it depends on $J_S$ and on the statistics of the single-particle spectrum and so
on magnetic field.

This is relevant for a quantitative understanding of the low-temperature behavior
of $\tilde{\alpha}\smeq1\smmi\langle G_{\text{peak}}\rangle_{\text{GOE}}/\langle
G_{\text{peak}}\rangle_{\text{GUE}}$ in closed QDs.\cite{FolkMH01} At $T\smeq0$,
since higher spin is more likely in the GOE case and since $\lambda$ is
smaller the bigger the spins involved, this renormalization leads to values
of $\tilde{\alpha}$ \textit{larger} than $0.25$---how much larger, of course,
depends on $J_S$.  At finite temperature, when several transitions contribute
to the conductance, it might also lead to values larger than $0.25$ and could
explain the small deviation observed at low temperature ($\kt\!\alt\!0.4\Delta$)
in Ref.  \onlinecite{FolkMH01}.  \footnote{Notice, however, that $\tilde{\alpha}$
always first decreases from its $T\smeq0$ value and then increases again.  This is
related to the fact that the contribution from the excited states come firts in
the GOE case.} Notice that either the CI model or dephasing processes lead to values
\textit{smaller} than $0.25$.\cite{BeenakkerSS01,EisenbergHA01,RuppAM02,HeldEA02}

\begin{figure}[t]
\begin{center}
 \includegraphics[width=8cm,clip] {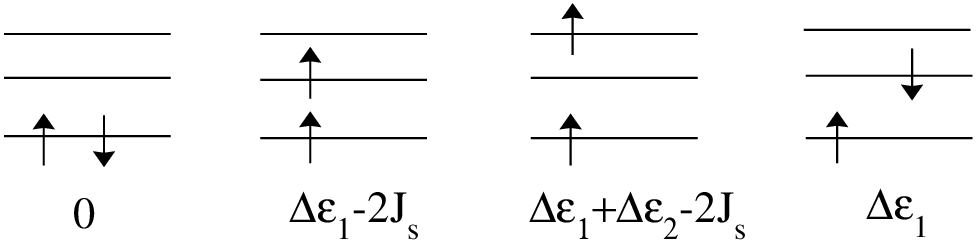}
\end{center}
\vspace{.5cm}
\begin{center}
 \includegraphics[width=8cm,clip] {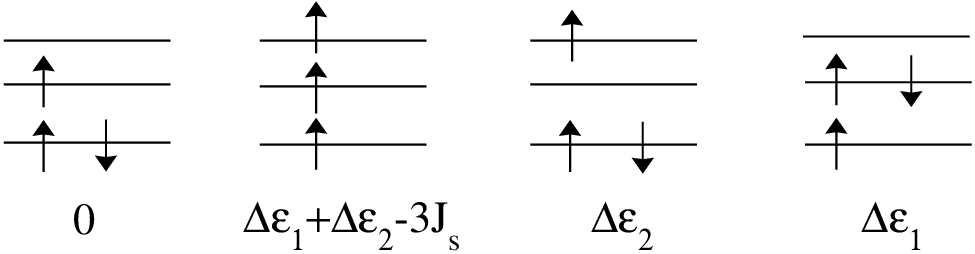}
\end{center}
\caption{Schematic representation of some low-energy states considered in the 
calculation of the PSD. The scheme at the top (bottom) corresponds to $N$ even 
(odd). 
Below each state is shown its energy difference from the left most one within the CEI 
model. $\Delta\ve_{1}$ and $\Delta\ve_{2}$ are the first and second level spacing 
respectively. 
}
\label{levels}
\end{figure}

%
%%%%%%%%%%%%%%%%%%%%%%%%%%%%%%%%%%%%%%%%%%%%%%%%%%%%%%%%%%%%%%%%%%%%%%%%%%
\subsection{Several state case}
\begin{figure}[t]
\begin{center}
 \includegraphics[width=8cm,clip] {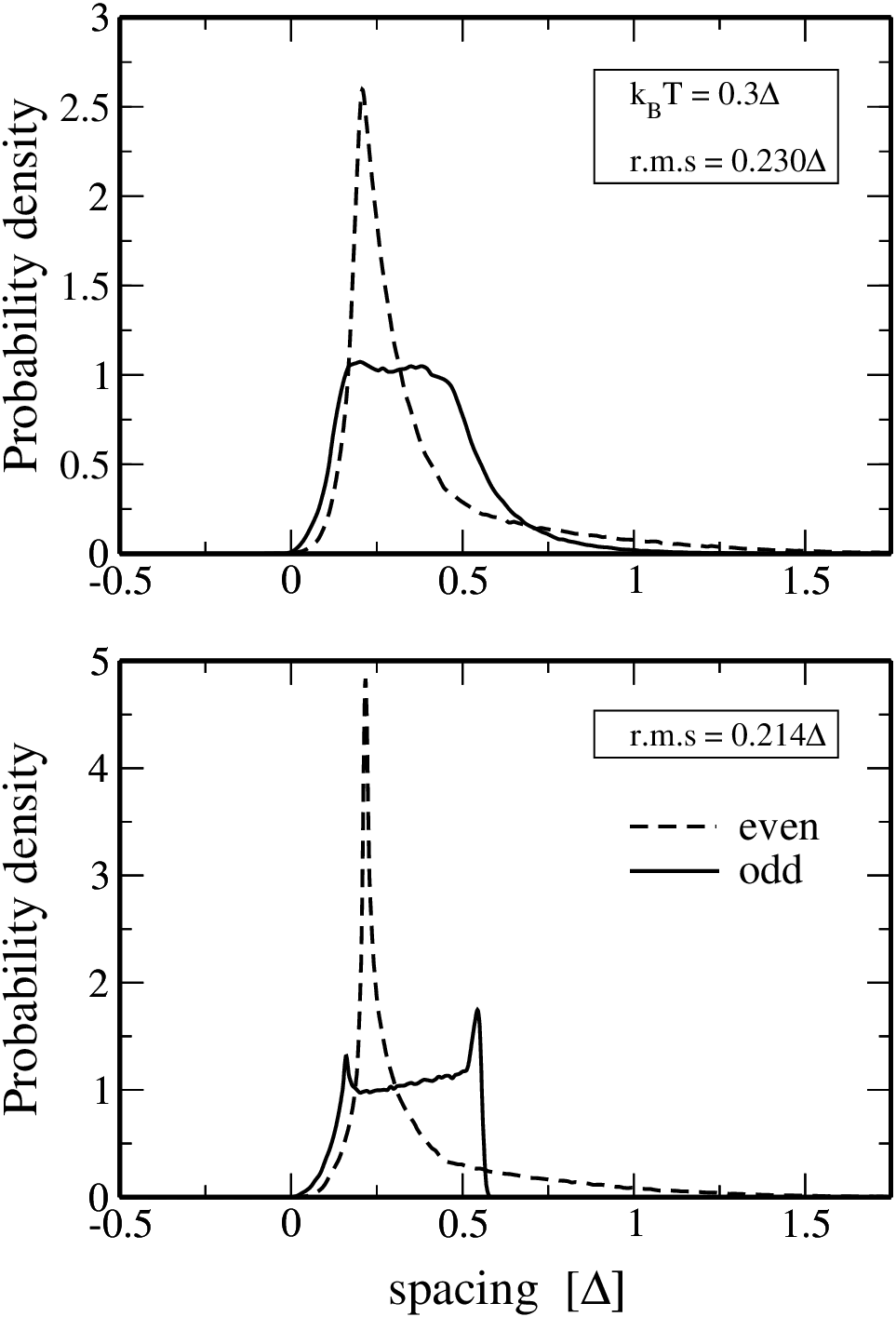}
\end{center}
\caption{Top panel: Finite temperature CB peak spacing distribution corresponding 
to the CEI model [Eq.~(\ref{CEI})] with 
$J_S \smeq 0.28\Delta$ ($r_s\smeq1$) and $\kt \smeq 0.3 \Delta$. Notice that the 
$\delta$-function in the odd distribution (solid line) is \textit{completely} 
smeared out by 
temperature and that the even distribution (dashed line) \textit{develops} a 
peak. 
Bottom panel: Same but neglecting the fluctuation of the coupling to the leads.  
Note that the long tail in the odd distribution (solid line) for large spacing is 
absent, and that the width of the peak in the even distribution is significantly 
reduced.  
}
\label{kTuH}
\end{figure}

\begin{figure}[t]
\begin{center}
 \includegraphics[width=8cm,clip] {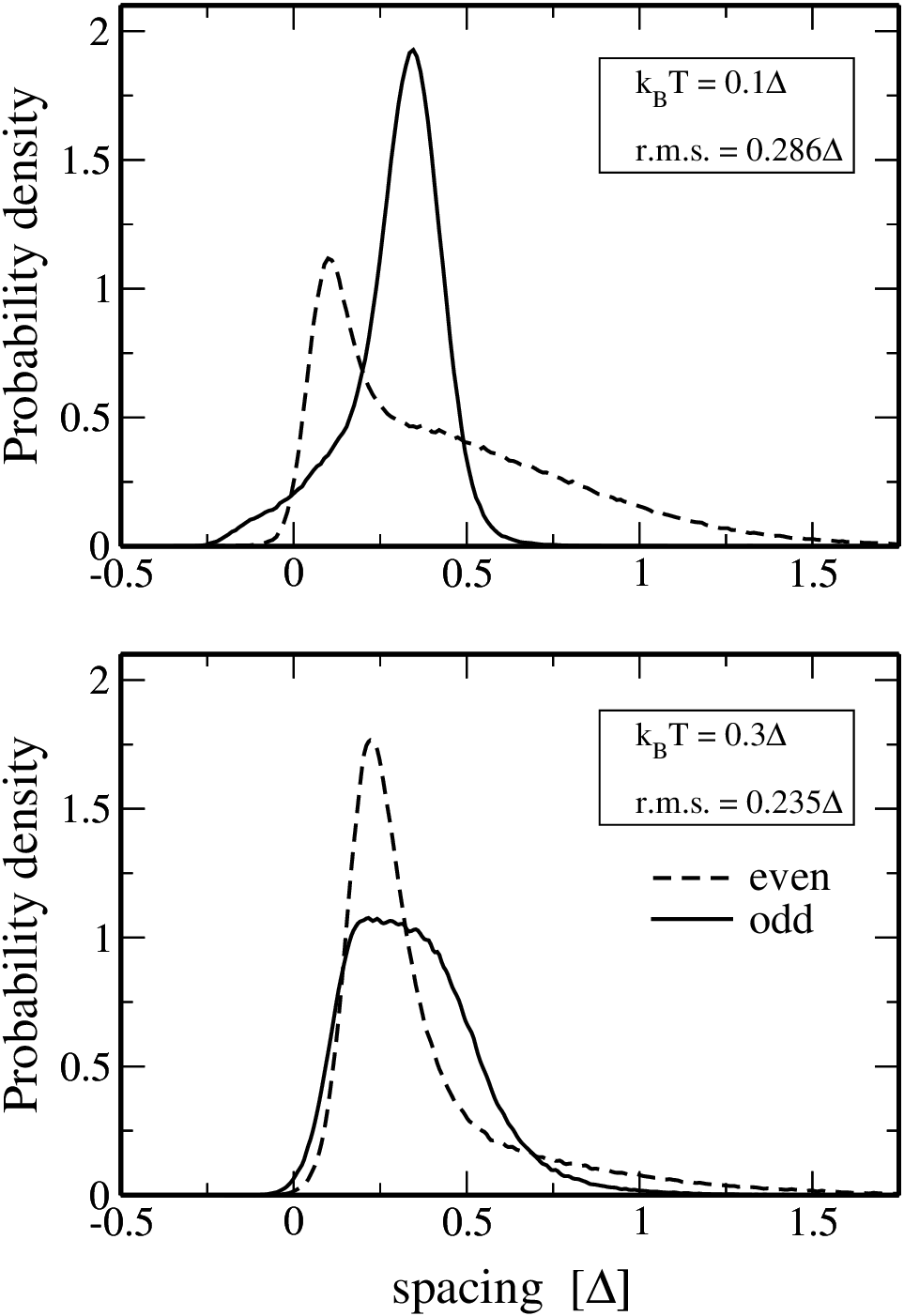}
\end{center}
\caption{Finite temperature CB peak spacing distribution corresponding to the 
Hamiltonian Eq. (\ref{leading}) with $J_s\smeq0.28\Delta$, $N\smeq500$, and
$b_{00}\smeq0.01$. The top (bottom) panel corresponds to $\kt\smeq0.1\Delta$ 
($0.3\Delta$) and the solid (dashed) line to $N$ odd (even)  
}
\label{kTPSD}
\end{figure}

%%%%%%%%%%%%%%%%%%%%%%%%%%%%%%%%%%%%%%%%%%%%%%%%%%%%%%%%%%%%%%%%%%%%%%%%%%
%
As we mentioned above, in the general case more than one transition contributes
to the conductance, and the CB peak position must be determined by maximizing
Eq.(\ref{G}) with respect to $\ag$. For arbitrary $T$, this requires the calculation
of all possible transitions between the eigenstates of $\Ha_{\text{QD}}$ with
$N\smmi1$ and $N$ electrons. For simplicity, we restrict ourselves to low enough
temperature so that only a few excited states are relevant.\footnote{Because the
fluctuations of the single particle levels, this is of course always violated for
some (very) rare realizations. This prevents us from describing the asymptotic
regime on the left side of the PSD in Fig. \ref{kTuH} and \ref{kTPSD}.} Therefore,
we kept $6$ states in the even case and $4$ in the odd one.  We checked numerically
that at $\kt\smeq0.3\Delta$ and for $J_S\smeq0.28\Delta$, the occupation of
these states is, on average, $99.4\%$ ($98.3\%$) for $N$ even (odd), being
smaller than $98\%$ ($92\%$) only about $1\%$ of the time.  In any case, the
effect of temperature can only be underestimated since, in general, different
transition leads to different spacing which in turns leads to an smearing of
the PSD.  Figure \ref{levels} shows some of the energy states considered in
the calculation of the PSD. For $N$ even, the lowest states with $S\smeq0$ and
$S\smeq1$ are the dominant states with an occupation of $52.4\%$ and $39.6\%$
respectively at $\kt\smeq0.3\Delta$ and $J_{S}\smeq0.28\Delta$.  In the odd case,
those are the lowest states with $S\smeq\frac{1}{2}$ and $S\smeq\frac{3}{2}$,
with $80\%$ and $7.5\%$ respectively.

The upper panel in Fig. \ref{kTuH} shows the CEI model PSD for non-zero temperature.
Besides the expected smearing of the sharp features and their shift due to the
entropic term in (\ref{cond}), there are two important new effects: (a) Temperature
alone is able to \textit{completely} wash-out the $\delta$-function, making the
odd distribution broader. Note in addition the long tail for large spacings;
we show below that the latter is \textit{not} simply thermal broadening.  (b)
The even distribution {\em develops} a peak at small spacings---in particular, the
maximum of the total distribution is dominated by the even distribution, in sharp
contrast to what occurs at $T\smeq0$. This strongly reduces the relative weight
of the long tail in the even case, and the distribution becomes less asymmetric.
Actually, the long tail is only slightly affected by temperature as it corresponds
to large values of the single-particle spacing.

The peak in the even distribution arises from cases where $S\smeq1$ and $S\smeq0$
states are  (almost) degenerate. It corresponds to the sharp discontinuity at
the origin in the $T\smeq0$ PSD---where both spin states significantly contribute
to the conductance.  Then, all the transitions with
 $\Delta\ve_1\!\simeq\!2J_S$ collapse into a single (average) value for the spacing,
which leads to a peak in the PSD.  According to Eq.~(\ref{gcond}) the corresponding
CB peaks are shifted by $\sim\!\pm\frac{1}{2}\kt\ln(4/2)$, which gives a total
shift of $\kt\ln2$ for the peak in the PSD.

Note that the r.m.s. of the distribution is reduced by temperature. In fact,
we found that it decreases monotonically from $T\smeq0$.

The fact that more than one transition contributes to the conductance implies that
the peak position also depends on the relative strength of the coupling to the leads
of the different levels ($\Gamma_{\alpha}$). This should be particularly important
when the GS and the first excited state are almost degenerate.  The bottom panel of
Fig. \ref{kTuH} shows the PSD assuming $\Gamma_{\alpha}\!\equiv\!\text{cte.}$ and
using the same parameters as before.  It is evident that much of the broadening
observed in the top panel is not directly caused by temperature but by the
\textit{fluctuation} of $\Gamma_{\alpha}$.

One of the most important differences is the absence of the long tail for large 
spacing in the odd distribution. This can be easily understood 
as follows.  First, let us note that the sharp jump in the $T\smeq0$ PSD at 
$J_{S}$ results from the transitions involving 
$S\smeq\frac{1}{2}$: $0\!\rightarrow\frac{1}{2}\!\rightarrow0$, 
$0\!\rightarrow\frac{1}{2}\!\rightarrow1$, 
$1\!\rightarrow\frac{1}{2}\!\rightarrow0$ and 
$1\!\rightarrow\frac{1}{2}\!\rightarrow1$. The spacing in each case is $J_{S}$, 
$\Delta\ve_{2}\smmi J_{S}$, $\Delta\ve_{1}\smmi J_{S}$ and 
$\Delta\ve_{1}\smpl\Delta\ve_{2}\smmi 3J_{S}$, respectively. It is easy to show 
that at $T\smeq0$, the conditions on $\Delta\ve_{1}$ and 
$\Delta\ve_{2}$ for each transition to occur, that is for the GS to have the 
appropriate spin, leads to a spacing $\le\!J_{S}$ in all the cases. 
This result is a consequence of the ``yes-no'' conditions required at $T\smeq0$.
At $T\!\neq\!0$, those conditions are relaxed and the last three transitions can
lead to a spacing bigger than $J_S$. This thermal broadening is responsible of
the disappearance of the $\delta$-function.  However, because the realizations
contributing to that part of the distribution have $\Delta\ve_{i}\simeq 2J_{S}$,
the thermal factors of the different transitions are very similar to each
other. Consequently, the relative strength of the couplings can overcome them:
the peak position is dominated by the most strongly coupled level, which might
correspond to the larger spacing. This explains the larger tail observed when
the fluctuation of $\Gamma_{\alpha}$ is taken into account.

Similarly, the width of the peak of the even distribution is strongly affected.
This clearly indicates that fluctuations of the wavefuntions of the QD 
\textit{strongly} modify the PSD. 

So far we have discussed temperature effects in the context of the CEI model. It 
is surprising that, even at this level of approximation, only a \textit{weak} 
even/odd effect or asymmetry is expected for $\kt\!\agt\!0.3\Delta$. 

%
%%%%%%%%%%%%%%%%%%%%%%%%%%%%%%%%%%%%%%%%%%%%%%%%%%%%%%%%%%%%%%%%%%%%%%%%%%
\subsection{Corrections to CEI model}
%%%%%%%%%%%%%%%%%%%%%%%%%%%%%%%%%%%%%%%%%%%%%%%%%%%%%%%%%%%%%%%%%%%%%%%%%%
%

Results including the leading order corrections are shown in Fig. \ref{kTPSD} for 
$\kt\smeq0.1\Delta$ and $0.3\Delta$ with the same parameters as in Fig. 
\ref{GSPSD}.
The additional fluctuations increase the broadening of the distribution. At low 
temperature, the effects of the residual interactions are dominant---though the 
$T$-induced-peak in the even distribution is evident. For $\kt\smeq0.3\Delta$, 
however, temperature is the main effect (compare with Fig. \ref{kTuH}). In this 
case, 
the even/odd effect is weaker but still noticeable---it should be kept in mind 
that the experimental noise may contribute significantly to weaken this effect. 
Also, notice that the PSD is \textit{not} Gaussian. A detailed 
analysis\cite{OngBHPM01} of the experimental data of Ref. 
\onlinecite{PatelCSHMDHCG98} shows that this 
is indeed the case. In fact, the agreement between these data and the PSD shown 
in the lower panel of Fig. \ref{kTPSD} is good, both qualitatively and 
quantitatively.

%%%%%%%%%%%%%%%%%%%%%%%%%%%%%%%%%%%%%%%%%%%%%%%%%%%%%%%%%%%%%%%%%%%%%%%%%
\section{Conclusions \label{conclu}}
%%%%%%%%%%%%%%%%%%%%%%%%%%%%%%%%%%%%%%%%%%%%%%%%%%%%%%%%%%%%%%%%%%%%%%%%%
In this work we have calculated the Coulomb blockade peak spacing distribution
including the most representative leading order corrections (up to
$\Delta/\sqrt{g}$) to the CEI model as well as the effect of finite temperature.

At $T\smeq0$, our results show that the PSD still presents a clear signature of an
even/odd effect.  Even though it is much weaker than the effect predicted by the
CI model, it is definitely big enough to be observable. No sharp features remain,
and the peak in the odd distribution (the former $\delta$-function) is still the
dominant characteristic.  Also, the distribution is far from being Gaussian and its
width is $\sim\!0.3\Delta$.  This number, however, depends on the value of $J_S$
and on many geometry-dependent parameters that could vary a bit for the actual QD.
On the other hand, it could be argued that the RPA approach used in the calculation
of the screening of the Coulomb potential is not appropriate for $r_s\!\sim\!1$.
We think, however, that the essential ingredients are captured by this approach
and that any correction that would arise from a more accurate calculation could be
included by a renormalization of $E_C$ and $J_S$. Such a renormalization could have
an important impact both in the shape of the distribution and in the strength of
even/odd effect since they are quite sensitive to the value of $J_S$.\cite{UllmoB01}
An {\it experimental} determination of $J_S$ should, then, be a high priority.

For $T\!\neq\!0$ the picture is quite different.  At $\kt\!\sim\!0.3\Delta$, the
roles of the even and odd distribution are inverted: it is the even distribution
that shows a peak, while in the odd one the $\delta$-function is washed-out. This
is very important, since the absence of the $\delta$-function has been one of
the puzzles in interpreting the experimental results.  The final distribution is
closer to a Gaussian-like shape, and the even/odd effect is much weaker. Here,
the main effect in the distribution comes from the temperature.  One important
consequence of the finite temperature is that the fluctuation of the coupling
to the leads becomes relevant and substantially contributes to the broadening of
the PSD.  Both the shape and the r.m.s. of the distribution agree with the data
in Ref.  \onlinecite{PatelCSHMDHCG98}.  We should mention however, that we fail
in reproducing the long tail for small spacing. This could be due to (a) relevance
of higher excited states---note that this part of the distribution corresponds to
cases where the single-electron spacing is very small---and the consequent effect
of the fluctuation of the couplings, or (b) mixing of the top levels caused by
the off-diagonal terms of the interaction---here our second order perturbation
theory fails---or by the off-diagonal terms of the scrambling.  At lower $T$
both distributions show a peak, which is a clear observable feature.

It is important to point out that our results do not explain the data of either Ref.
\onlinecite{SimmelAWKK99} or \onlinecite{LuscherHEWB01}.  Nevertheless, in both
cases there are some elements to think that this is not a ``failure" of the model.
In Ref. \onlinecite{SimmelAWKK99} the interpretation of the transport process
itself is not clear.\cite{AbuschMagderSWKK00} For example, the width of the CB
peak is not controlled by temperature though its shape corresponds to a thermally
assisted process.  In Ref. \onlinecite{LuscherHEWB01} temperature effects are
negligible due to both $\kt\!\sim\!0.05\Delta$ and $J_S\!\sim\!0.25\Delta$.
However, the single-particle dynamics is not fully chaotic because of the regular
shape of the QD. Therefore, the effect of regular orbits and of the presence of
regular islands in phase space, must be considered. 
This could enhance the contribution of $\Pi_\text{B}$ in Eqs. (\ref{varchi}) and 
(\ref{varM}) and lead to larger fluctuations of both $M_{\alpha,\beta}$ and 
$\mathcal{X}_{\alpha,\beta}$. This subject is quite complex
and we leave it for future work.  Nevertheless it is important to mention that
the r.m.s. in Ref.  \onlinecite{LuscherHEWB01} is of order $0.4\Delta$ and that
there is a weak even/ odd effect.

\begin{acknowledgments}
We thank D. Ullmo for many valuable conversations and appreciate helpful 
discussions with I. L. Aleiner, L. I. Glazman,  and Ph. Jacquod.
GU acknowledges partial support from CONICET (Argentina). This work was supported 
in part by the NSF (DMR-0103003).
\end{acknowledgments}

%%%%%%%%%%%%%%%%%%%%%%%%%%%%%%%%%%%%%%%%%%%%%%%%%%%%%%%%%%%%%%%%%%%%%%%%%
\appendix
%%%%%%%%%%%%%%%%%%%%%%%%%%%%%%%%%%%%%%%%%%%%%%%%%%%%%%%%%%%%%%%%%%%%%%%%%

%%%%%%%%%%%%%%%%%%%%%%%%%%%%%%%%%%%%%%%%%%%%%%%%%%%%%%%%%%%%%%%%%%%%%%%%%
\section{The diffusive case}
%%%%%%%%%%%%%%%%%%%%%%%%%%%%%%%%%%%%%%%%%%%%%%%%%%%%%%%%%%%%%%%%%%%%%%%%%

Similar results for the leading order corrections can be obtained in the case of 
diffusive QDs after the appropriate change of some definitions.\cite{AleinerBG02} 
First, the dimensionless conductance is now 
given by $g\smeq\hbar \gamma_{1}/\Delta$ where $\gamma_{1}$ is the smallest 
non-zero eigenvalue of the diffusion equation, 
$\smmi D\nabla^{2}f_{n}(\br)\smeq \gamma_n\, f_{n}(\br)$, supplemented with von 
Neumann boundary conditions. 
$D\smeq v_{\text{F}}\ell/2$ is the diffusion constant and $\ell$ the mean free 
path. Second, 
$k(\br_1,\br_2)\smeq\exp(\smmi r/\ell) J_0^{2}(\kf r)$ with 
$r\smeq|\br_1\smmi\br_2|$. Third, the propagator $\Pi_{\text{B}}(\br_1,\br_2)$ is 
replaced by its diffusive counterpart 
\be
\Pi_{\text{D}}(\br_1,\br_2)\smeq\frac{\Delta A}{\pi} \sum_{n} \frac{f_{n}(\br_1) 
f_{n}(\br_2)}{\hbar\gamma_{n}}.
\ee
In order to obtain a numerical value for the fluctuation of the different matrix 
elements, a specific geometry must be assumed. For a disc of radius $R$ we get,
\be
\gamma_{m,n}\smeq\frac{D\beta^2_{m,n}}{R^2}, \qquad J_{m}'(\beta_{m,n})\smeq0
\ee 
and
\be
f_{m,n}(\br)\smeq A_{m,n}\left\{{\cos m\phi \atop \sin 
m\phi}\right\}J_{m}^2(\beta_{m,n}\frac{r}{R})
\ee
with 
\be
A_{m,n}\smeq 
\sqrt{\frac{2}{(1\smpl\delta_{m,0})\pi}}\frac{\beta_{m,n}}{R\sqrt{\beta_{m,n}^2\smmi 
m^2}|J_{m}(\beta_{m,n})|}.
\ee
Here, $J_{m}(x)$ is a Bessel function and $J'_{m}(x)$ its derivative. Note
that the first non-zero eigenvalue, $\gamma_{1,1}$ is proportional to the
square of the first zero of $J'_1$, $\beta_{1,1}\smeq1.84$. This means that
the last mode to relax is the one with the smallest number of nodes in the
angular direction and none in the radial direction (except for the origin),
$f_{1,1}(\br)\!\propto\!\cos{\phi}\, J_{1}^{2}(\beta_{1,1}r/R)$.  Then, $g\smeq\hbar
D\beta_{1,1}^{2}/R^2\Delta\!\propto\!\ell/R\sqrt{N}$. For $N\!\sim\!500$ and
$\ell\!\sim\!R/2$ this gives $g\!\approx\!13$.

%%%%%%%%%%%%%%%%%%%%%%%%%%%%%%%%%%%%%%%%%%%%%%%%%%%%%%%%%%
\subsection{Scrambling}
%%%%%%%%%%%%%%%%%%%%%%%%%%%%%%%%%%%%%%%%%%%%%%%%%%%%%%%%%%

The main contribution in this case comes from the second term in Eq.  (\ref{varchi}). 
Because there is no reason to assume a $1/r$ decay of the wavefunction 
correlation in a general case, we cannot use the same approach we used in 
Section \ref{evalvarchi}. Instead, we assume a disc geometry to get\cite{AleinerBG02}
\bea
\nonumber
b_{00}&\smeq&\frac{1}{\pi 
A}\sum_{\gamma_{0,n}>0}{\frac{\gamma_{1,1}}{\gamma_{0,n}}\left[\int \text{d}\br 
V(\br)\, f_{0,n}(\br)\right]^2}\\
\nonumber
	&\smeq&\frac{1}{4\pi}\sum_{\beta_{0,n}>0}{\frac{\beta_{1,1}^2}{\beta_{0,n}^4 
}\frac{\sin^2(\beta_{0,n})}{J_{0}^2(\beta_{0,n})}}\\
&\simeq&0.004
\eea
which is very close to the value obtained for the ballistic case. 
Here we used $V(\br)\smeq1/4[1\smmi(r/R)^2]^{\frac{1}{2}}$.
Thus
$\text{var}(\mathcal{X}_{\alpha,\beta})$ remains of the same order as 
in the ballistic case. 

%%%%%%%%%%%%%%%%%%%%%%%%%%%%%%%%%%%%%%%%%%%%%%%%%%%%%%%%%%
\subsection{Diagonal elements}
%%%%%%%%%%%%%%%%%%%%%%%%%%%%%%%%%%%%%%%%%%%%%%%%%%%%%%%%%%

In this case, the terms in Eq. (\ref{varM}) that involve $k(\br_1\smmi\br_2)$  
are small (assuming $\ell\!\ll\!\sqrt{A}$) and can be neglected. 
Then\cite{AleinerBG02,AlhassidG01}  
\be
\text{var}(M_{\alpha,\beta})\smeq\frac{\Delta^2}{4A^2}\int\text{d}\br_1\text{d}\br_2\, 
\left[\Pi_{\text{D}}(\br_1,\br_2)\right]^{2}.
\label{varMapp}
\ee
Assuming a disc geometry we get $\text{var}(M_{\alpha,\beta})\smeq c_2 
\Delta^2/4g^2$ with 
\bea
\nonumber
c_2&\smeq&\frac{1}{\pi^2}\sum_{\gamma_{m,n}}\left(\frac{\gamma_{1,1}}{\gamma_{m,n}}\right)^2\\
\nonumber
&\smeq&\frac{2}{\pi^2}\sum_{m=0}^{\infty}\sum_{n=1}^{\infty}\left(\frac{\beta_{1,1}}{\beta_{m,n}}\right)^4\\
&\simeq&0.13
\eea
Once again, the numerical result is similar to the one we obtained for the 
ballistic case. For instance, using $N\!\sim\!500$ and $\ell\!\sim\!R/2$, we get 
$\text{rms}(M_{\alpha,\beta})\!\approx\!0.014\Delta$, which should be compared 
with the second row in Table \ref{table1}. 

%%%%%%%%%%%%%%%%%%%%%%%%%%%%%%%%%%%%%%%%%%%%%%%%%%%%%%%%%%%
\section{Second order correction for $S\smeq\frac{1}{2}$}
%%%%%%%%%%%%%%%%%%%%%%%%%%%%%%%%%%%%%%%%%%%%%%%%%%%%%%%%%%%

We give here an explicit expression for the second order correction to the energy 
due to the off-diagonal terms  for the case of  $S\smeq\frac{1}{2}$,
\begin{widetext}
\bea
\nonumber
E^{(2)}_{S=\frac{1}{2}}&\smeq&
    \sum_{\beta\geq\alpha,\gamma\geq\delta}{\frac{3\left| 
u_{\alpha,\beta,\gamma,\delta}\right|^{2}
                \smpl\left| 
a_{\alpha,\beta,\gamma,\delta}\right|^{2}}{\ve_{\alpha}\smpl\ve_{\beta}\smmi\ve_{\gamma}\smmi\ve_{\delta}}
(1\smmi\frac{\delta_{\alpha,\beta}}{2})(1\smmi\frac{\delta_{\gamma,\delta}}{2})}
      \smpl \sum_{\alpha,\gamma\geq\delta}{\frac{3\left| 
u_{\alpha,1,\gamma,\delta}\right|^{2}
                \smpl\left| 
a_{\alpha,1,\gamma,\delta}\right|^{2}}{2(\ve_{\alpha}\smpl\ve_{1}\smmi\ve_{\gamma}\smmi\ve_{\delta})}
                 (1\smmi\frac{\delta_{\gamma,\delta}}{2})}\\
\nonumber
&&      \smpl \sum_{\beta\geq\alpha,\gamma}{\frac{3\left| 
u_{\alpha,\beta,\gamma,1}\right|^{2}
                \smpl\left| 
a_{\alpha,\beta,\gamma,1}\right|^{2}}{2(\ve_{\alpha}\smpl\ve_{\beta}\smmi\ve_{\gamma}\smmi\ve_{1})}
                 (1\smmi\frac{\delta_{\alpha,\beta}}{2})}
\smpl\sum_{\beta,\gamma}{\frac{\left| \sum_{\alpha\neq\beta}' (3 
u_{\alpha,\beta,\gamma,\alpha}\smpl a_{\alpha,\beta,\gamma,\alpha})
		(1-\frac{\delta_{\alpha,1}}{2})\right|^{2}\smpl\frac{3}{4}\left| 
u_{1,\beta,1,\gamma}\smpl a_{1,\beta,1,\gamma}\right|^{2}}
				{2(\ve_{\beta}\smmi\ve_{\gamma})}}\\
&&\smpl\sum_{\beta}{\frac{\left| \sum_{\alpha\neq\beta} (3 
u_{\alpha,\beta,1,\alpha}\smpl a_{\alpha,\beta,1,\alpha})
		\smpl a_{\beta,1,1,1}\smpl a_{1,1,1,\beta}\right|^{2}}	
{16(\ve_{\beta}\smmi\ve_{1})}}
\smpl\sum_{\gamma}{\frac{\left| \sum_{\alpha} (3 u_{\alpha,1,\gamma,\alpha}\smpl 
a_{\alpha,1,\gamma,\alpha})\right|^{2}}
				{4(\ve_{1}\smmi\ve_{\gamma})}}
\eea
\end{widetext}
where $\alpha$ and $\beta$ ($\gamma$ and $\delta$) correspond to doubly occupied 
(empty) levels of the $S\smeq\frac{1}{2}$ state---except for the term with 
$\sum'$ 
in which case the singly occupied level labeled ``$1$'' is included. Note that 
there are several terms which add coherently.
% that were neglected in Ref. \onlinecite{JacquodS01}.

%%%%%%%%%%%%%%%%%%%%%%%%%%%%%%%%%%%%%%%%%%%%%%%%%%%%%%%%%%%%
\bibliographystyle{apsrev}
\bibliography{usaj}
\end{document}